\documentclass[sigconf,nonacm]{acmart}
\AtBeginDocument{%
  }

\begin{document}

\title [Dynamic Theater] {Dynamic Theater: Location-Based Immersive Dance Theater, Investigating User Guidance and Experience}

\author{You-Jin Kim}
\email{yujnkm@ucsb.edu}
\orcid{0000-0003-0903-8999}
\affiliation{%
  \institution{University of California,}
  \city{Santa Barbara}
  \state{}
  \country{USA}
}

\author{Joshua Lu}
\email{joshualu@berkeley.edu}
\affiliation{%
  \institution{University of California,}
  \city{Berkeley}
  \state{}
  \country{USA}
}

\author{Tobias Höllerer}
\email{holl@cs.ucsb.edu}
\affiliation{%
  \institution{University of California,}
  \city{Santa Barbara}
  \state{}
  \country{USA}
}

\renewcommand{\shortauthors}{Kim et al.}

\begin{teaserfigure}
  \includegraphics[width=\textwidth]{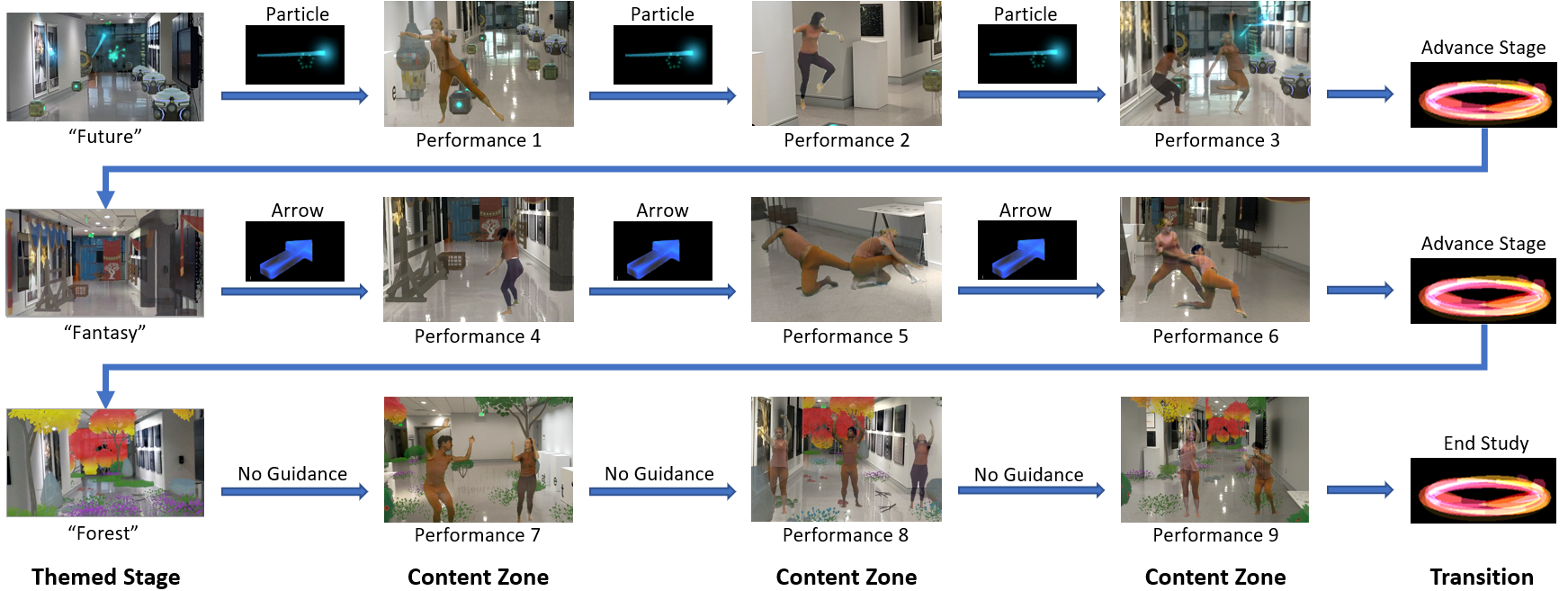}
  \caption{Overview of the user experience trajectory. Within one physical space, users maneuver through three themed stages aided by a guidance system (``Particle” or ``Arrow”) for navigation during dance performances. Users are directed to the ``Content Zone” for virtual dance performances.}
  \label{fig:teaser}
 \Description[User experience trajectory]{An overview of the user experience trajectory in the Dynamic Theater.}
\end{teaserfigure}

\begin{abstract}
  Dynamic Theater explores the use of augmented reality (AR) in immersive theater as a platform for digital dance performances. The project presents a locomotion-based experience that allows for full spatial exploration. A large indoor AR theater space was designed to allow users to freely explore the augmented environment. The curated wide-area experience employs various guidance mechanisms to direct users to the main content zones. Results from our 20-person user study show how users experience the performance piece while using a guidance system. The importance of stage layout, guidance system, and dancer placement in immersive theater experiences are highlighted as they cater to user preferences while enhancing the overall reception of digital content in wide-area AR. Observations after working with dancers and choreographers, as well as their experience and feedback are also discussed.

  \smallskip
\textit{ This is a preprint version of this article. The final version of this paper can be found in the Proceedings of ACM VRST 2023. For citation, please refer to the published version.}
\textit{This work was initially made available on the author's personal website [yujnkm.com] in September 2023, and was subsequently uploaded to arXiv for broader accessibility.}
\end{abstract}

\ccsdesc[500]{Human-centered computing~Empirical studies in HCI}
\ccsdesc[500]{Computing methodologies~Mixed / augmented reality}
\ccsdesc[500]{Computing methodologies~Perception}

\keywords{\keywords{Mobile Augmented Reality, Wide-Area, User Study, Immersive Theater}}


\maketitle

\section{Introduction}
Theater choreographer McAuley emphasizes the intricate interplay between space, performer, and narrative. Choreographers must consider the theater building's overall physical arrangement when planning dance sequences. McAuley suggests that creative teams should collectively explore the building and stage before dance rehearsals to enhance engagement~\cite{mcauley1999space}.

Dance floorwork encompasses movement throughout the entire stage, emphasizing the role of expression and movement in human bodies~\cite{diana2018overlooked}. In contrast, digital media consumption typically involves minimal physical activity. While modern gaming can offer expansive virtual environments, recent advancements in Extended Reality (XR) aim to replicate aspects of live theater performances~\cite{erkert2003harnessing, lella2014us, cheok2002interactive, gochfeld2018holojam, oculus2017dear, tender2019under}.

Previous studies demonstrated the potential of presenting virtual content through augmented reality (AR) in large-scale settings~\cite{feiner1997touring,thomas2002first,cheok2004human,rompapas2019towards, kim2022investigating, kumaran2023impact}. They employ multiple tracking and registration mechanisms such as tracking and occlusion models, cloud anchor points, and user-centric design elements to ensure comfort and safety during navigation. Current XR practices, however, still underutilize available space. To optimize the utilization of a large indoor space for the application domain of immersive dance theater, we investigate user preferences on guidance systems to enhance the theatrical experience for a mobile AR audience. Focusing on choreography, which encompasses the entire floor area and involves user interaction, we explore how the effective utilization of space can showcase immersive dance pieces.

\begin{figure}[tb]
 \centering 
 \includegraphics[width=\columnwidth]{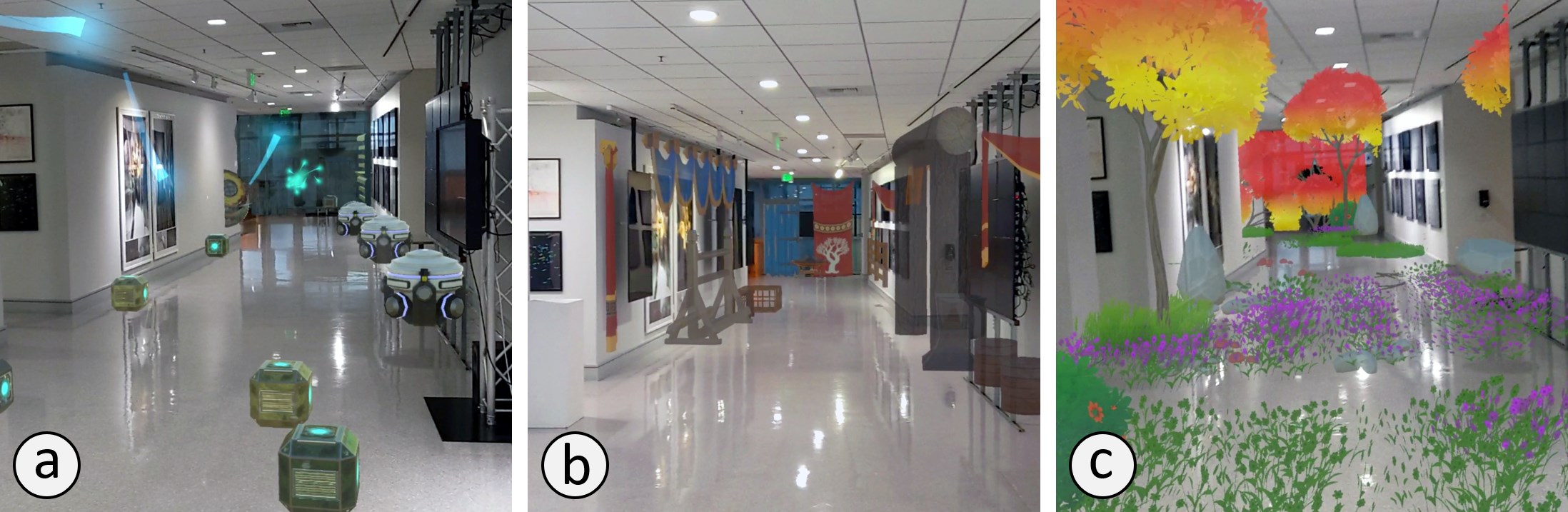}
 \caption{Three virtual stages were captured from the headset using Microsoft MR Capture. Here, both virtual objects and the physical environment are visible within the (a) future stage, (b) fantasy stage, and (c) forest stage.}
 \label{fig:all-scenes}
\end{figure}

According to McAuley, theater intertwines fact and fiction, creating a nuanced and cohesive theatrical experience~\cite{mcauley1999space}. In the context of AR theater attendance, our aim is to allow users to navigate the narrative at their own pace. Therefore, a guidance system that subtly enhances the core experience while seamlessly blending into the environment is necessary.

Furthermore, the placement of digital content in a vast open area poses a significant challenge in facilitating users' intended discovery of AR narrative content~\cite{cheok2002interactive, windschitl2000virtual, winn2001learning}. Ensuring freedom to navigate the room while implementing direction to specific physical locations for viewing digital content requires meticulous consideration of space arrangement.

Location-based narratives in open areas enable natural locomotion, which replicates real-world exploration and allows users to navigate at their own pace. Previous studies showed a preference for realistic modes of travel~\cite{sayyad2020walking}. Leveraging location-based narratives embraces the benefits of natural locomotion while addressing associated challenges. We build upon the foundation of interactive theater experiences in virtual environments on performance stages~\cite{gochfeld2018holojam}, expanding it to various location-based indoor environments. Our work, Dynamic Theater, offers curated immersive experiences, going beyond traditional in-person theater by enabling user interaction with visual content and exploration of the environment.

\begin{figure}[tb]
 \centering 
 \includegraphics[width=\columnwidth]{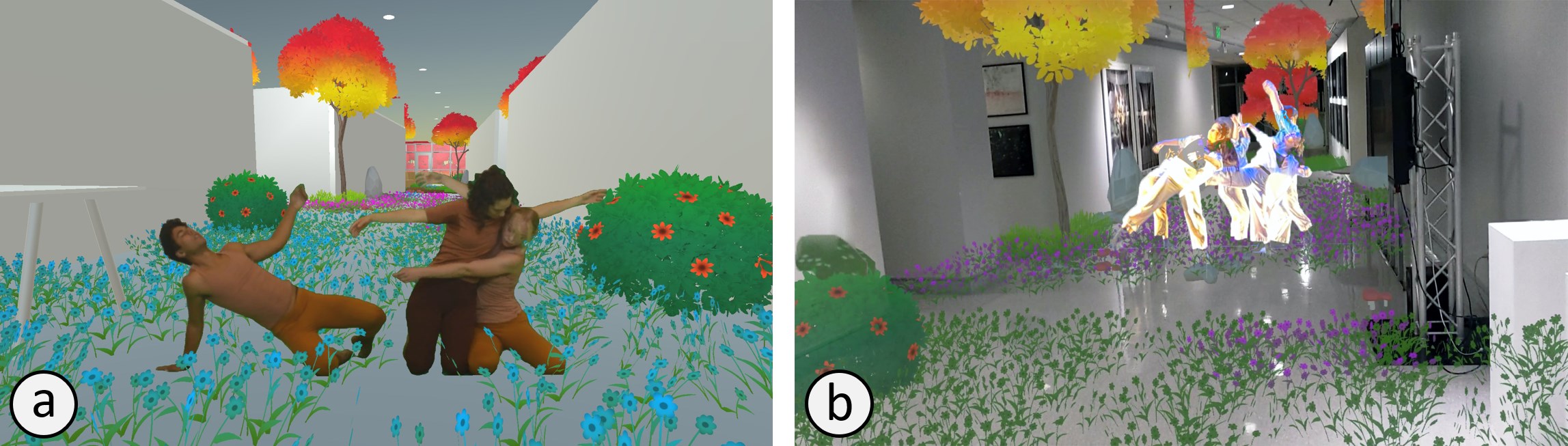}
 \caption{A digital twin of the physical space was used to situate the dancers correctly in relation to the environment to preserve their footwork. Image (a) features an illustration shown to the choreographer for precise dancer placement in Unity while (b) shows how dancers were projected through the headset.} 
 \label{fig:billboarding}
\end{figure}

Dynamic Theater presents an immersive system showcasing recorded dance performances in a large indoor space. The goal is to create an augmented dance experience supported by a visually complementary and unobtrusive user guidance system, experienced through the HoloLens 2. Using billboarding techniques, transparent dancer holograms are strategically positioned in collaboration with a professional choreographer to integrate dance movements into the physical arrangement (see Figure \ref{fig:billboarding}). Navigational guidance systems based on virtual particle systems are evaluated to enable a narrative-driven AR theater experience. Through the deployment of various guidance methods and themed virtual stages, insights are obtained towards optimizing future stagecraft and dynamic user experiences. Our contributions to the immersive theater experience include the following:

\begin{itemize}
    \item Developed a Dynamic Theater system to seamlessly merge dance choreography with physical surroundings. Performance playback involves carefully placed billboarded projections of recorded dancers, enabling anyone with the headset to enjoy the play within the designated area at any time.
    \item Created a narrative-driven AR theater experience utilizing navigational guidance systems with projected content. This facilitates natural audience locomotion, guiding viewers to content zones. 
    \item Extracted valuable insights from surveys and interviews with 6 dancers and a choreographer. Their input, especially regarding dancer involvement, influenced the project direction throughout.
\end{itemize}

    Altogether, our work enables the expansion of interactive theater to diverse location-based AR theater design, involving AR content tailoring to physical layouts and collaborations with dancers for floor and footwork integration.

\section{Related Work}

In this section, we review augmented reality (AR) research on storytelling, interactive design, and location-based narratives.

\subsection{Storytelling in Mixed Reality}


A shift in recent years from traditional theater production to mixed reality theater production, which shares many of the same components, presents new opportunities and fresh obstacles\cite{coulombe2021virtual, gochfeld2022tale}. Mixed reality has been extensively studied for narrative presentation, allowing users to experience curated stories in immersive environments that enhance the conveyance of artistic works \cite{geigel2004theatrical}. Various performances, including choreographed performances, have been adapted to mixed reality, such as \emph{The Life} \cite{abramovic2020life}, \emph{Debussy3.0} \cite{clay2014integrating}, and \emph{Nautilus} \cite{fischer2016nautilus}, integrating digital elements and augmented visual effects into live dance performances. Other works involve close collaboration between the artist and technical engineers to create augmented dance experiences \cite{clay2012interactions}. Our project builds upon these prior works, extending the foundations of choreography in mixed reality to create a more interactive, and mobile, user experience. 

While our focus is on dance performances, insights from other narrative theater experiences have also informed our research. Works like \emph{CAVE} \cite{layng2019cave} and \emph{CAVRN} \cite{herscher2019cavrn} offer shared cinematic experiences in large virtual environments for collective audiences. \emph{Gulliver} \cite{ARGulliver} is a play that incorporates AR headsets and augmented effects, while the use of physical props (e.g., Augmented Playbill, Prayer Wheel, and Tarot Cards) expands the theatrical experience beyond the stage \cite{nicholas2021expanding}. Building upon the insights gained from narrative theater experiences, our research aims to further explore the potential of mixed reality in the context of dance performances.

\subsection{Interactive Design with Locomotion}


Our understanding of human decision-making and time perception in mixed reality experiences is growing rapidly~\cite{awe, FLAVIAN2019547}. Immersion enhances focus and extends perceived time, potentially leading to increased satisfaction. An element of awe can expand time perception, influence decision-making, and enhance well-being. The perception of self within the play relates to the interactive narrative experience~\cite{self}. 

Research has explored locomotion, particularly walking, as a fundamental input for spatial computing in mixed reality designs, especially in location-based elements of AR games. This approach allows users to physically explore the environment and engage with interactive narratives. While games and performances differ, the concept of creating location-based experiences aligns closely with our research. Notably, \emph{Soul Hunter} \cite{weng2011soul} demonstrates how players navigate multiple rooms in a castle and interact with translucent ghosts using an AR weapon. Other AR games, such as \emph{ARQuake} \cite{thomas2000arquake} and \emph{HoloRoyale} \cite{rompapas2018holoroyale}, aimed to bring first-person shooter experiences to larger physical spaces. Additionally, Madsen et al. \cite{madsen2022fear} explore integrating an exploration-based horror game into an indoor AR environment. Collaborative and multi-user experiences are facilitated through AR applications such as \emph{Human Pacman}\cite{cheok2004human}, where users interact as characters in a large physical environment, and \emph{MapLens} \cite{morrison2009like}, which encourages collaborative task-solving in a city-wide AR setting. Notably, the combination of natural walking experiences and sightseeing has been used to highlight and preserve important historical events using a mixed reality interactive narrative approach \cite{lehto2020augmented, fujihata2022behere, wither2010westwood}. Our research expands the understanding of mixed reality in theater by exploring the possibilities and challenges of integrating dance performances into designated spaces while maintaining a strong connection with the audience and the surrounding environment through the interplay of narrative, locomotion, and location.

\subsection{Location-Based Narrative}


In location-based works, the relationship between the narrative and physical setting is essential for enhancing the immersive and realistic aspects of the storytelling \cite{location}. In this section, we discuss works that extend beyond dancing, focusing instead on the utilization of location-based storytelling and interactive narrative as a means to convey stories \cite{han2023architectural}.

As briefly mentioned, sites of historical significance rely on the integration of physical locations with digital information to deliver impactful experiences. For example, \emph{BeHere 1942} transports users back in time to the actual location where the forced expulsion of Japanese Americans occurred at the old Santa Fe train station in LA \cite{fujihata2022behere}. \emph{Lights On!} is an augmented reality game that connects users to real cultural heritage sites \cite{lehto2020augmented}. \emph{The Westwood Experience} narrates the story of a mayor and guides users to physical locations mentioned in the narrative itself \cite{wither2010westwood}. Furthermore, an AR application developed for the Oakland Cemetery guides users through the physical cemetery using audio narrations and stories of its former inhabitants \cite{dow2005exploring}.

Mixed reality theaters have also embraced innovative approaches to creating interactive experiences. Cheok et al. combined location-based, outdoor exploration with a virtual theater performance of a Shakespearean play. Users are guided through an outdoor environment with content zones that provide clues, eventually transitioning into an indoor virtual reality environment featuring a captured actor portraying Hamlet \cite{cheok2002interactive}. Other mixed reality theaters offer opportunities for users to actively participate in the theater performance itself \cite{gochfeld2018holojam, pietroszek2022meeting, pietroszek2022dill, lyons2023gumball}. For instance, \emph{Holojam in Wonderland} enables audience members to follow a live actor in the form of an avatar, engaging with both the actor and the virtual environment on stage \cite{gochfeld2018holojam}. \emph{Story CreatAR} helps facilitate the process of incorporating locomotive storytelling in AR, particularly when narrative elements can occur at various physical locations within large environments \cite{singh2021story}.

In addition, recent work explored a novel approach to adapting narratives to real spaces for AR experiences \cite{li2023locationaware}. This system utilizes an optimization-based algorithm that automatically assigns contextually compatible locations to story events. Other techniques offer the audience more choices in terms of where they can experience the same content, including location-specific places, such as museums, or an alternative VR environment \cite{geigel2020digital}. Our approach promotes active decision-making in integrating digital content into live performances and physical spaces, curated by artists.

\begin{figure}[t]
\centering \includegraphics[width=0.5\columnwidth]{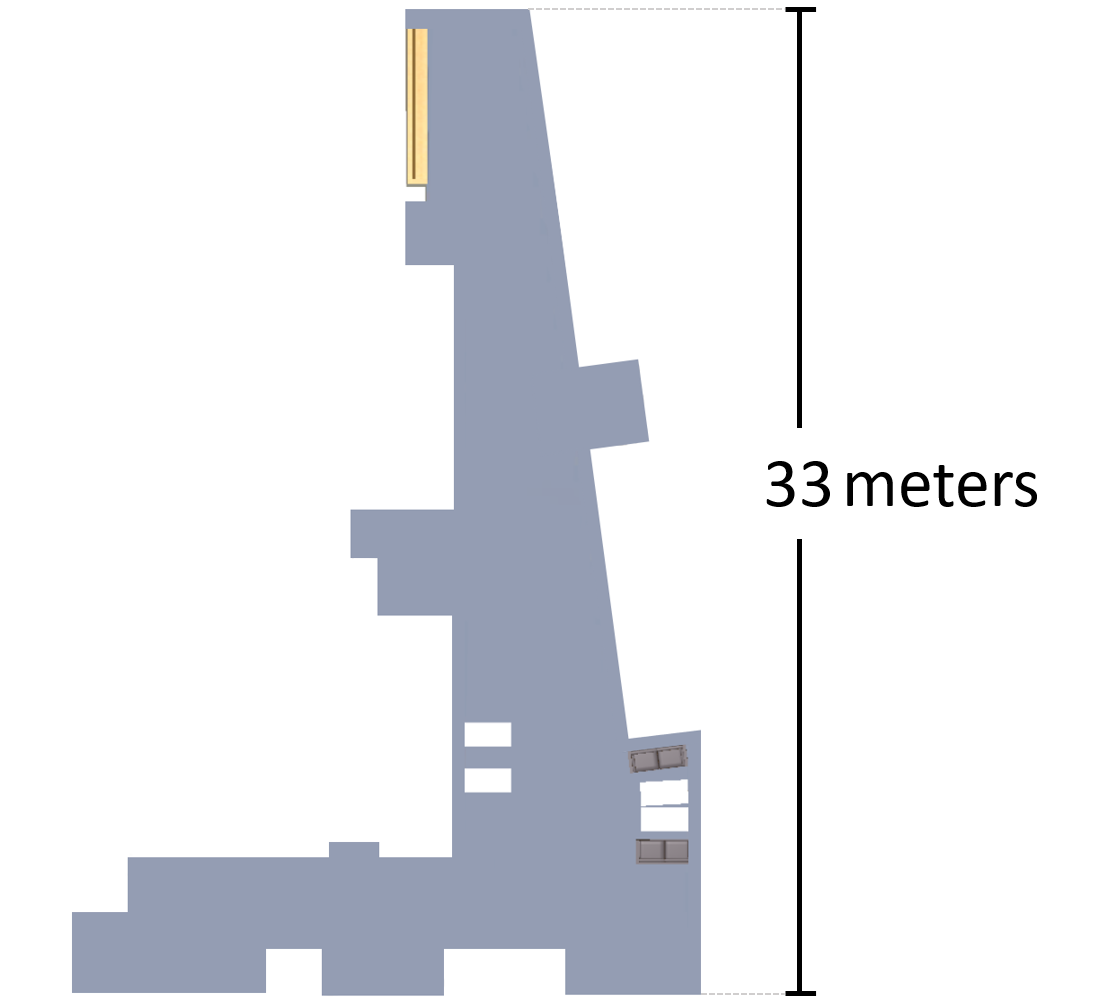}
 \caption{Top view of the experiment area and its physical layout. The dimensions are 208.54$m^{2}$ (2,244 $ft^{2}$).}
 \label{fig:top-view}
 \end{figure}

\section{Dynamic Theater}
During the development of the Dynamic Theater system, we encountered technical challenges in tracking, floor plan modeling for the occlusion layer, and anchoring strategies. Collaboration with artists, dancers, and choreographers, thus, was crucial. This section explores the procedure of capturing dancer movements and presenting them in the HoloLens 2. We also discuss the design of the Virtual Stage, the guidance system, and the user experience.

The theater piece took place in a spacious indoor environment with two hallways for user traversal (Figure \ref{fig:top-view}). The indoor space measured approximately 208.54 $m^{2}$. Virtual objects strategically restricted access to certain areas. Three distinct virtual stage designs showcased the versatility of the space.

\subsection{Capturing Dancer Movements}
We collaborated closely with six professional dancers and one choreographer to capture dance movements for the Dynamic Theater. Before the dance session, dancers tried out the headset and experienced the virtual stages in the physical location firsthand. Performances took place on a green-screen stage built specifically for this project. Dancers did not wear additional technology; instead, we used 5 depth cameras, 2 DSLR cameras, and 2 action cameras. In the following sections, we elaborate on equipment used, and the dance capture process.

\begin{figure}[tb]
 \centering 
 \includegraphics[width=\columnwidth]{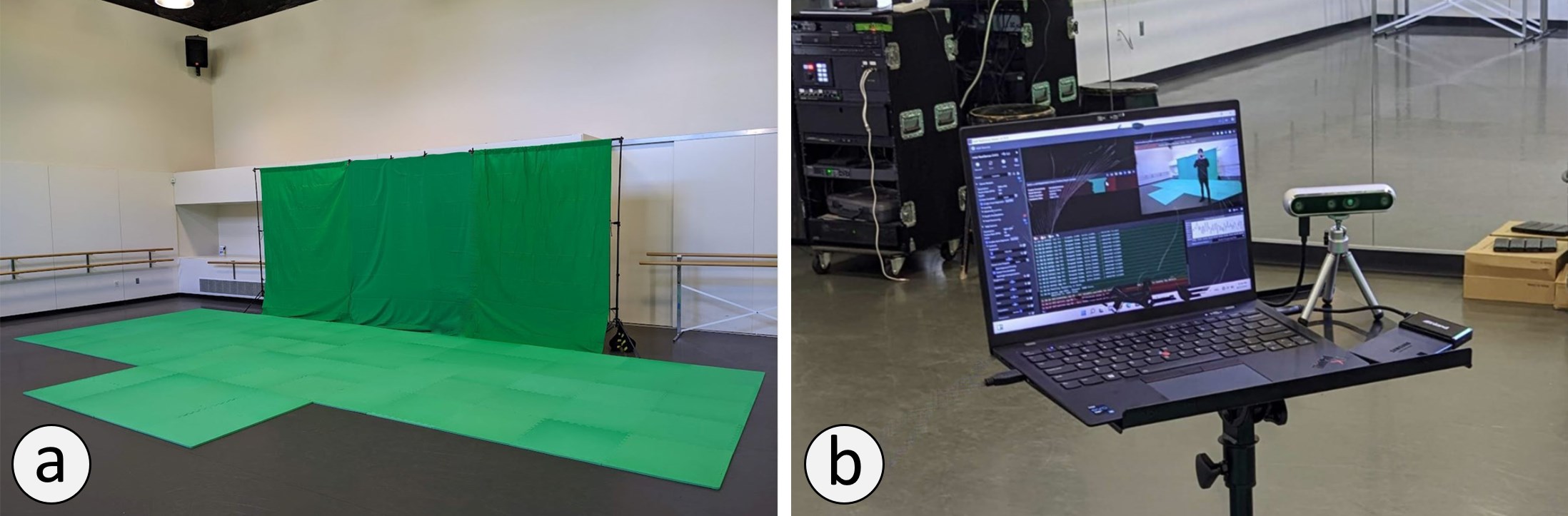}
 \caption{(a) The dance capture session employed a green screen setup on a stage measuring 2.9 $m$ x 14.6 $m$, with a green screen backdrop measuring 11 $m$ x 2.6 $m$. (b) Five RealSense D455 cameras, along with a laptop and a 1.2 $m$ tall stand, were positioned between the three DSLR cameras to capture depth video frame data.}
 \label{fig:green-screen}
\end{figure}

\subsubsection{Equipment and Green Screen Setup}
The dance capture process spanned 5 days, including setup, testing, and 2 days of recording with all dancers and the choreographer. Setup took 6 hours, with 3 graduate student helpers present. They operated cameras and ensured dancer safety by maintaining the cleanliness of the green screen dance floor mat.

The dance studio was equipped with 85-inch TV screens on each wall and a sound system, which played music and showcased the virtual stage during the sessions.

For video capture, we used 3 Nikon D7500 DSLR cameras. One camera was positioned at the center, 7.5 $m$ away from the screen, with a wide-angle 24 $mm$ lens to capture the entire width of the green screen. The other 2 cameras were placed at a 45-degree angle to capture different perspectives. Additionally, 5 Intel RealSense D455 depth cameras were positioned between the DSLRs but closer to the screen, approximately 5.5 $m$ away.

The dance stage featured 1.27 $cm$ thick green dance mats securely affixed using flooring adhesives and sealers. The stage measured 2.9 $m$ in width and 14.6 $m$ in length, with an additional 11 $m$ extension of the green screen backdrop. The dance mats extended 1.8 $m$ beyond each end of the screen to accommodate user footwork preparation. The screen, measuring 2.6 $m$ in height, was safely fastened to the wall structure of the dance studio.

\subsubsection{Dance Capture Session}
The dance capture session and user study received the relevant ethics commission and human subject approval. Permission was obtained from the university's dance department to conduct the project. A Zoom meeting was held among the authors, tech specialists, helping crews, 6 dancers, and a choreographer to discuss dance moves, dress code, the story's overall arc, and the theme of the virtual stage. The meeting clarified the data collection process and the intended use of the captured video and data, while respecting the dancers' opinions. The choreographer and authors reviewed the story line, cue sheets, and shared their vision, considering the green screen setup possibilities. The dance capture sessions took place over 2 consecutive weekdays from 9 am to noon, totaling 6 hours. Sessions included warm-up, stretching, a review of the story board and cue sheets, and rotating dancers in groups of different sizes. Finally, consent and talent release forms were collected along with an exit interview.

\begin{figure}[t]
\centering \includegraphics[width=\columnwidth]{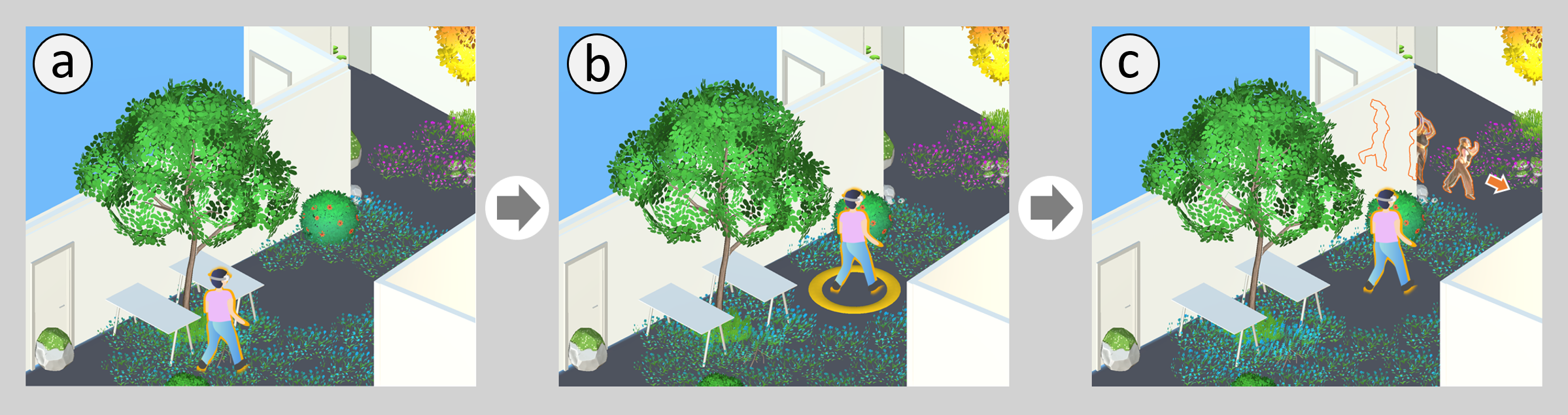}
 \caption{The location-based trigger allowed the viewer to experience the performance as the choreographer intended. The location-based trigger remains invisible and users are unaware of its presence while navigating (a). Upon entering the predetermined trigger spot (b), the dancers emerge (c), often from behind the walls, revealing themselves to the viewers.}
 \label{fig:locationtrigger}
 \end{figure}

\subsection{Digital Content Placement}
To enable active decision-making in dancer placement, virtual stage prompts, and location-based event triggers, an accurate floor model and a digital twin of the designated AR play area were needed. The digital twin was created using the Matterport 3D camera system, Apple's LiDAR scanning, Blender 3D, and the Unity game engine. This digital twin served multiple purposes, including visualizing the virtual stage in conjunction with projected objects and handling occlusions on the HoloLens device. Discussions with the choreographer were conducted to envision dance moves and determine dancer placement, considering the physical layout and content simultaneously. This approach allowed for strategic placement of content zones, location triggers, and stage advance triggers along the users' natural walking path. The nine dance content zones were positioned with distances ranging from 10 to 15 $m$ apart to facilitate exploration and navigation.

Azure Spatial Anchors were utilized to achieve precise content alignment during startup on our HoloLens 2 devices~\cite{buck2022azure}. The saved spatial anchors from the server were automatically loaded to position digital content, in turn handling occlusions caused by physical walls.

Location-based triggers, invisible to the user, were strategically positioned along pathways with a radius of 1$m$. These triggers activate corresponding content when the user enters the designated collision zone, resulting in the deployment of dance performances. Dancers fade in and can be concealed behind a wall when utilizing the building structure.

\begin{figure}[t]
\centering \includegraphics[width=\columnwidth]{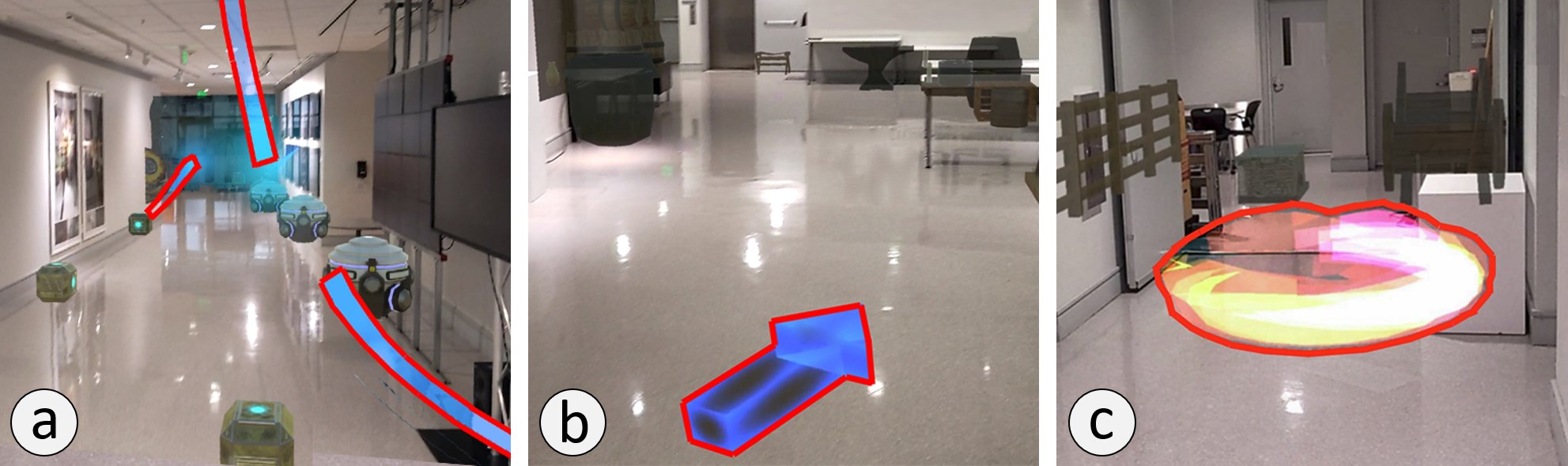}
 \caption{Visual display of our three types of navigational aids. (a) The particles: a long, visible streak of light is emitted by each particle for better view. Particles move towards the general direction of the target, with slight variations in their trajectories (b) The arrow: a typical and straightforward guidance mechanism. (c) The spiral: serves as a transition, advancing users to the next stage or ending the study.}
 \label{fig:guidance-systems}
 \Description[Three types of navigational aids]{Three types of navigational aids. Shows ``Particles'', ``Arrow'' and ``Spiral'' navigational aids.}
 \end{figure}

\subsubsection{Dancers Placement}
To achieve volumetric reconstruction of dancers, we initially intended to use our RealSense depth video frame data. However, this approach proved to be computationally intensive for the HoloLens 2 due to its hardware limitations, such as the Qualcomm Snapdragon 850 mobile processor and limited accessible RAM (2 GB) for external applications. Considering our goal of including detailed accessories and multiple sets of dancers on the virtual stage, which required loading full physical models, animated objects, and music, we needed to manage the additional computational costs. As a result, we opted for a 2D billboarding technique to simulate the three-dimensionality of the dance performance. This involved rendering captured dancers, a volumetric virtual stage, dynamic particle systems, and other interactive components in 3D, creating the illusion of depth when combined. Spatial audio, another crucial element for perceived 3D depth, was implemented using a separate audio source with the built-in audio spatializer, enhancing users' spatial attention in the content-rich environment.

While considering the use of avatars and skeletal animation keyframes derived from video frame data using body tracking, we prioritized maintaining the visibility of the actual dancers. Therefore, we decided to utilize 4K video footage captured from the DSLR cameras.

The captured videos were processed to extract the dancers by employing background removal techniques such as luma keys and opacity masks. This resulted in the creation of videos with a transparent background. To preserve transparency, the edited videos were exported in the {\small WebM} format, which supports an alpha channel. Within our development pipeline, these videos were integrated into materials using a fade blending mode, opaque render type, and a d3d11 vertex shader to produce the intended hologram effect.

\begin{figure}[t]
 \includegraphics[width=\columnwidth]{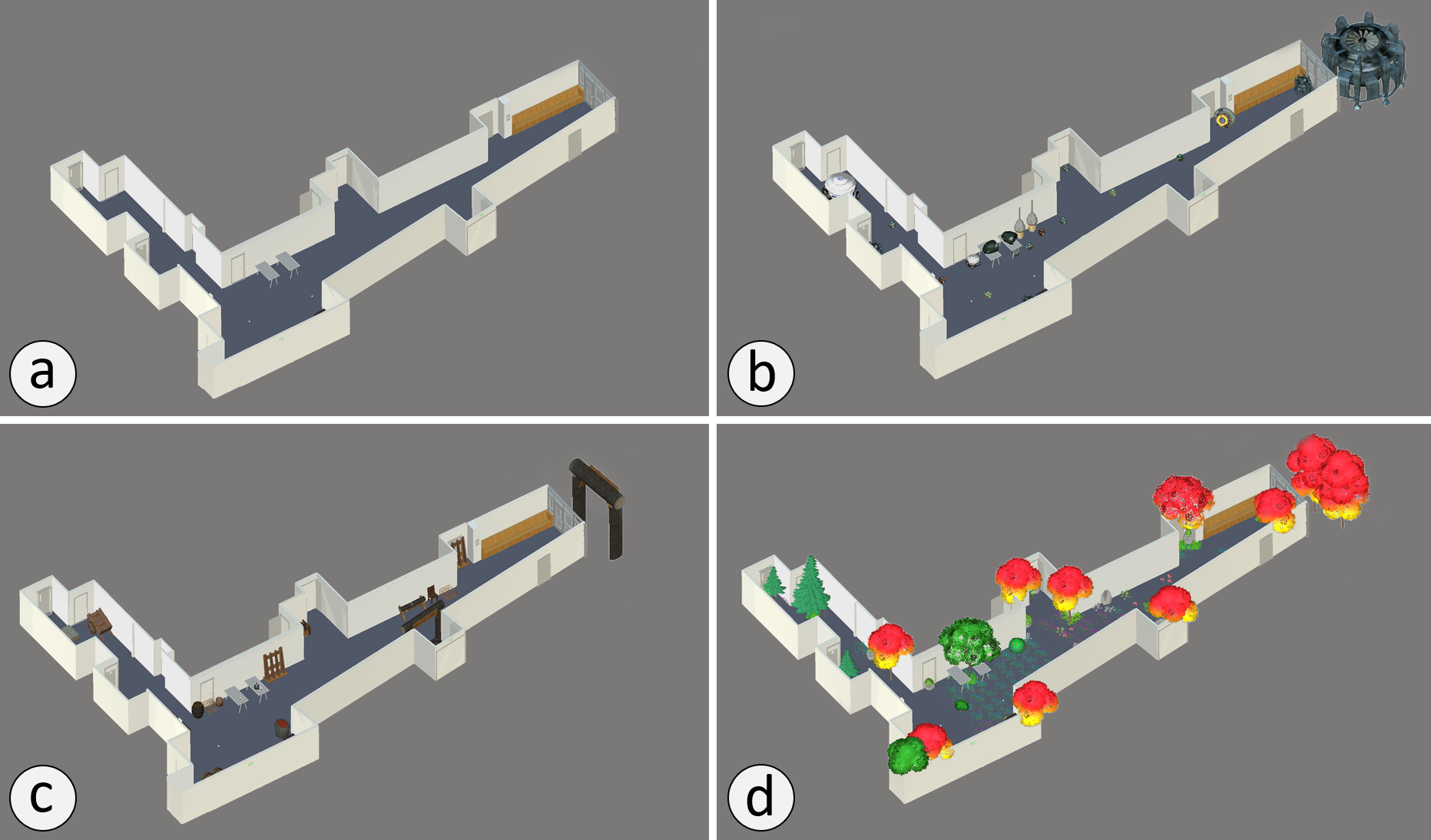}
 \caption{This figure shows an aerial view of (a) the digital twin of the physical layout and the physical layouts of (b) the ``Future”, (c) ``Fantasy”, and (d) ``Forest” stage prompts.}
 \label{fig:all-scenes}
 \end{figure}

\subsection{Guidance Systems}
To provide effective and non-intrusive guidance during the interactive experience, we implemented two types of guidance systems: a particle system and a traditional arrow system. These systems directed users to the content zones, blending with the overall experience and enhancing user engagement. Our goal was to offer subtle cues without rushing the user, allowing them to fully enjoy the surrounding content. Both the particle system and arrow system were used as the primary navigational aid, guiding users towards the dancers or the Stage Advancement Trigger.

\subsubsection{Particle Guidance}
Particle systems have enhanced the integration of visual effects in theatrical performances, particularly in choreography. For instance, Visual Flow effectively combines particle effects with dance performances, eliminating the requirement for dancers to synchronize with pre-recorded effects \cite{brockhoeft2016interactive}. Our system takes a similar approach, combining particles with dancers in a digital theater within an indoor physical environment not specifically designed for performances. Users observe one to six guidance particles gracefully floating towards the target location, appearing at the outer periphery of their vision (Figure \ref{fig:guidance-systems}a). This provides a steady flow of indicators without overwhelming the screen. To ensure clear directionality, a colored streak marks the particles' trail, extending only a short distance to avoid clutter. Randomness is applied to the particle movement, simulating the appearance of naturally glowing insects or sparks to blend with the environment. The particles' spawn point is also randomized within the user's head position which introduces slight variations in their path towards the target. To counteract any deviation caused by accumulated noise, the velocity direction of the particles is reset to point directly towards the target every second.

\subsubsection{Arrow Guidance}
The second guidance system we implemented is the arrow guidance system (Figure \ref{fig:guidance-systems}b), which uses a simple mechanism of pointing towards the next content zone. The arrow remains positioned 40$cm$ above the ground and 2$m$ away from the user's head, appearing at the center of the user's field of view (FOV) within the HoloLens display. The blue 3D arrow, measuring 30$cm$ in length and 10$cm$ in width, indicates the direction in which the user should walk. As the user reaches the content zone and a dancer appears, the arrow gradually fades away until the dance performance is completed.

\subsubsection{Other Spatial Guidance}

In addition to the arrow guidance system, we enhanced the user's guidance through other spatial cues. A spatial audio system played louder music from the content zone, creating an audio cue to direct the user's attention. Furthermore, a red light spinning spiral (Figure \ref{fig:guidance-systems}c) is positioned at the desired location, serving as a marker for the user to move to in order to transition to the next scene. This spiral system remains invisible and inactive until all the dance sequences within the content zone have been observed, ensuring that users do not transition prematurely. It functions as a cue, similar to the closing of a theater curtain before revealing a new stage.

\subsection{User Experience Trajectory}
The play consists of three themed virtual stages: Future, Fantasy, and Forest, each with three content zones featuring 10- to 30-second dance sequences. Our rationale for choosing these vastly distinct virtual stage environments was to illustrate the diverse experiences permitted in the same physical space. The design aims for a 7-minute experience, with users spending 2-3 minutes on each stage. Guidance is provided in two stages, while one stage is navigated without guidance. The order of presentation was Future, Fantasy, and Forest, with varied guidance system. After watching all three dance performances in the final stage (Forest), the spiral advance location trigger appears, concluding the trial and user study. The Interactive Narrative, displayed in Figure \ref{fig:teaser}, includes spiral indicators for stage advancement and play conclusion.

Each stage was comprised of three choreographed performance segments. Users activated each segment individually by following guidance indicators. When a performance concludes, the spiral system appears, and the guidance indicators shift focus to guide the user towards it.

\begin{figure*}[t]
 \includegraphics[width=\textwidth]{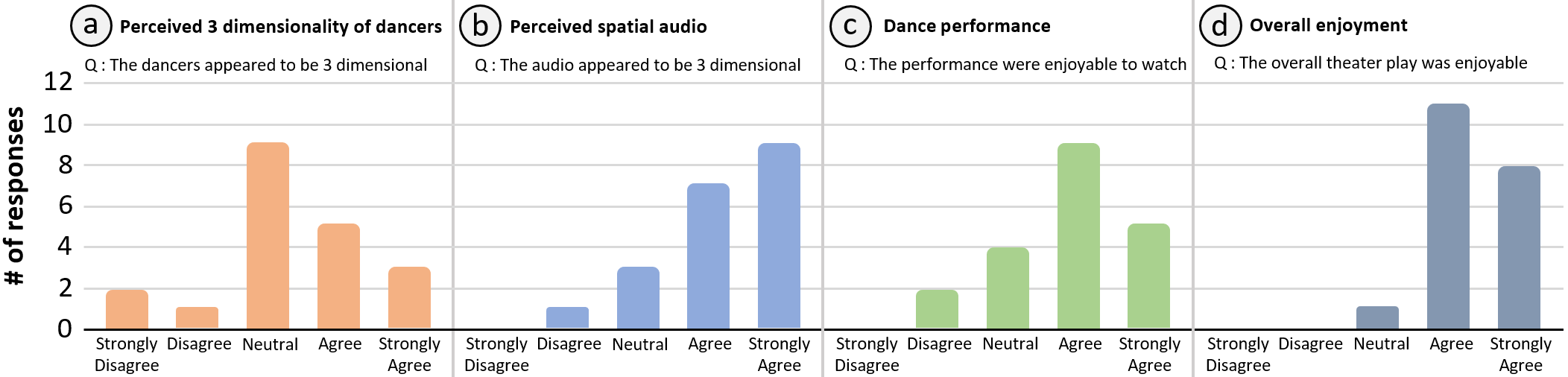}
 \caption{User responses to questions regarding their evaluation of the environment on a five-point Likert scale (1 = Strongly disagree, 5 = Strongly agree). (a) Q: ``The dancers appeared to be 3D.'' The mean response was 3.3. (b) Q: ``The audio appeared to be 3D.'' The mean response was 4.2. (c) Q: ``The performances were enjoyable to watch.'' The mean response was 3.85. (d) Q: ``The overall theater play was enjoyable.'' The mean response was 4.35.}
 \label{fig:user-eval}
 \end{figure*}

\begin{figure}[tb]
 \centering 
 \includegraphics[width=\columnwidth]{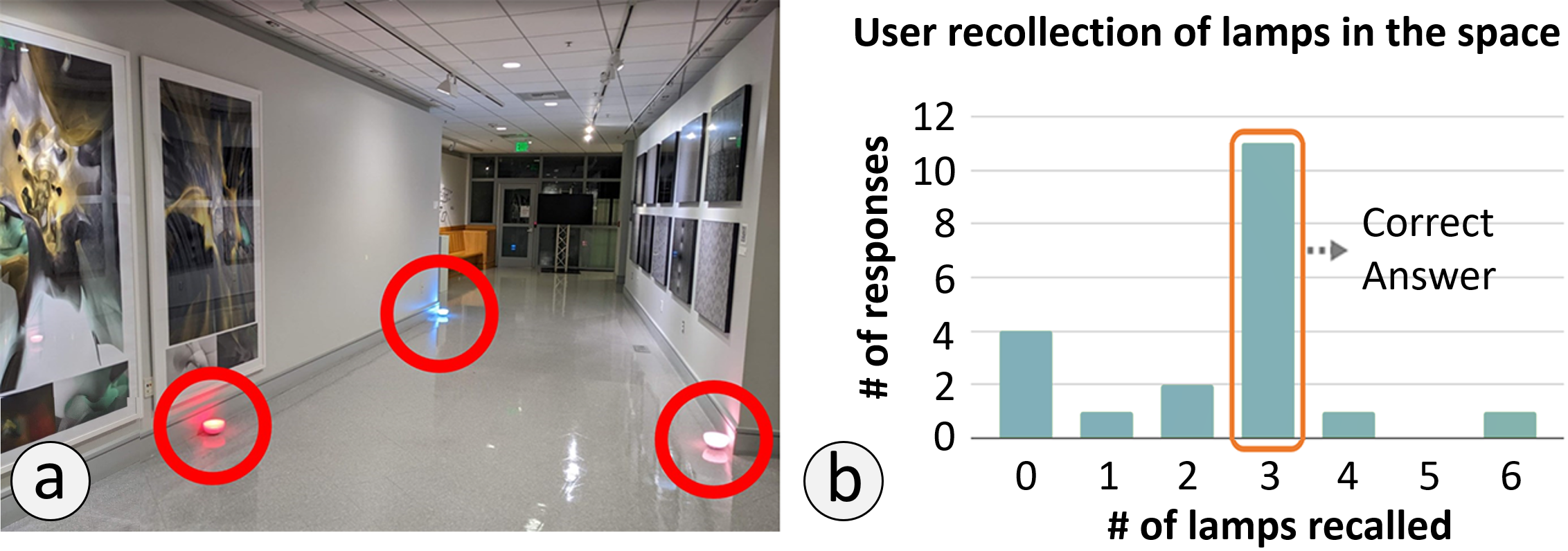}
 \caption{(a) A view of the physical environment with the three lamps on the ground. For each user, lamps were arranged and placed in random locations. These lamps were used to test users' ability to recall the physical layout, as the users were not given any instruction regarding the floor lamps. (b) Bar graph of users’ answers to how many lamps they noticed on the ground without having been alerted to them beforehand. The average answer is 2.4, with 55\% recalling the correct number (3).}
 \label{fig:lamp-setup}
\end{figure}

\section{Experiment}
This experiment examined user navigation and completion time in an AR theater play. Feedback and insights from the experiment improved the user experience of narrative interaction. Drawing on previous work in interactive narrative and AR theater projects, which explored innovative ways to interact with extended reality, as well as users' responses to spatial arrangements, we formulated the following hypotheses:
\begin{itemize} 

    \item H1: The Dynamic Theater provides an enjoyable experience and creates the perception of an extended passage of time.
    \item H2: The guidance system facilitates users' natural locomotion and content viewing without compromising the AR experience.
    \item H3: The particle guidance system in the Dynamic Theater outperforms a traditional arrow as a guidance mechanism, offering effective navigation and visual appeal.

\end{itemize}

\subsection{Design}

Users explored all three virtual stages using both guidance systems (particles and arrows) in a counterbalanced manner. To mitigate the order effect, users were randomly assigned labels A or B, in equal proportion. Label A included a future stage with a particle indicator, a fantasy stage with an arrow indicator, and a forest stage with no indicator. Label B had a future stage with an arrow indicator, a fantasy stage with a particle indicator, and a forest stage with no indicator. The trial proceeded without intervention to measure the play's length. On average, users completed the experience in 5 minutes and 5 seconds (SD = 48.27 seconds), with each stage designed for approximately 100 seconds. Typically, users took 4 to 7 minutes to finish the entire play. The walking distance per stage was approximately 53 m, with a total distance of around 150 meters for the entire play.

\subsection{Apparatus}
The experiment took place in a 208.54 $m^{2}$ (2,244 $ft^{2}$) indoor space, where participants were instructed to stay within a designated area. Compliance with this rule was observed. Three Philips Hue Go Bluetooth Controllable Portable LED lamps, with a diameter of 15 $cm$, were arranged in a circular pattern. The Microsoft HoloLens-2 Mixed Reality headset was used, employing low-polygon-count 3D models with varying levels of detail to manage computational load. The HoloLens utilized single-pass instanced rendering and vertex-lit shading techniques.

When designing the virtual stages, we followed guidelines to create distinct experiences that differed from the physical space. Each stage consisted of approximately fifty items, including large, medium, and small objects. The goal was to cover around 20\% of the environment in each stage with projected prompts, ensuring that the entire wall was not completely covered.

\subsection{Questionnaire}
Dancers completed a comprehensive questionnaire consisting of four sections and 25 questions (11 Likert-scale items and 14 open-ended responses). The questionnaire covered topics such as the green screen setup, dance schedule, assistive media and technology, and instruction. To explore the audience reception of the AR experience, a separate user study was conducted with 20 participants, involving 27 questions (17 Likert-scale items and 10 short answers), to gather insights.

\subsection{Participants}

For the audience study, we recruited 20 adult participants (9 men, 11 women), ranging in age from 18 to 34. Six participants used vision correction, and all reported normal or corrected-to-normal vision. 15 participants had no prior experience with augmented reality. All participants reported normal walking abilities and were able to walk normally during the trial.

\subsection{Procedure}

Upon arrival, participants received a guided tour of the walking area, emphasizing the virtual fence location for safety. Informed consent forms were signed, and a pre-questionnaire capturing demographic information was completed. Participants were informed about the guidance indicators leading to dance performance segments and encouraged to walk continuously at a normal pace. Portable LED lamps emitting distinct colors had been placed randomly along the hallway but were not mentioned.

The HoloLens 2 was activated, and Azure Spatial Anchors ensured accurate alignment between the virtual scene and physical environment. Participants wore the headset while a researcher remained nearby for safety and observation. 

At the end of the performance, an exit interview and questionnaire were conducted to gather feedback and insights. No instances of discomfort or dizziness were reported.
Participants completed a questionnaire assessing the guidance system's visual and functional qualities and their perception of scene duration. Additional questions covered the dance performance and participants' recall of the floor lamps. The entire procedure lasted approximately 25 minutes per participant.

\section{Results}
Data was collected on dance performance, perceived play time, user ratings of the AR experience, and guidance systems. Feedback from the dancers and the choreographer was also gathered. Paired t-tests and ANOVA tests were conducted to compare the results between the different guidance indicator types using a significance level of $\alpha=0.05$.

\subsection{User Evaluation of the Dynamic Theater}
Figure \ref{fig:user-eval} presents plots of user responses on a 5-point Likert scale regarding their experience of viewing the AR dance theater performance. The question about the visual appeal of our virtual dancers yielded a mean response of 3.85, with 70\% of respondents agreeing or strongly agreeing that the performance was interesting and enjoyable to watch (see Figure \ref{fig:user-eval}c). The mean response to the question \textit{``The dancers appeared 3D''} was 3.3 (Figure \ref{fig:user-eval}a), and for \textit{``The audio source seemed 3D''}, the mean response was 4.2 (Figure \ref{fig:user-eval}b).

The overall rating users assigned to the theater experience was 4.35 (\ref{fig:user-eval}d). One user expressed, \textit{``This experience has proven to be surprisingly refreshing. The innovative approach, to me, ranks much higher than existing and conventional performances.''} Another user's favorite aspect was \textit{``exploring a virtual world while having the real world as a palette.''}

The project was well received by users, who voiced their individual tastes and enjoyment of various features of the play, which aligned with the Dynamic Theater project's concept of catering to individual experiences without rushing. As one user stated, \textit{``I enjoyed walking through the forest scene due to the lack of restrictions.''} Another user found joy in \textit{``going into the teleportation spiral,''} despite its relatively smaller role in the overall experience. The combination of virtual flowers and dancers left a lasting impression, with one user noting, \textit{``Seeing the virtual flowers and dancers together was something else, it was great.''} The dynamic nature of the music, changing with the dancers' entrance into scenes, was particularly enjoyable for one user, while another appreciated \textit{``the music and its ability to change based on location.''} The simple act of following the particles provided a sense of exploration, and the familiarity of virtual stages in known places was an intriguing aspect noted by one user. Additionally, the use of stereo audio added a layer of fun and immersion, as described by a user who delighted in \textit{``spinning my head listening to the sound move''} and likening it to \textit{``listening to 3D audio.''}

\begin{figure}[tb]
 \centering 
 \includegraphics[width=\columnwidth]{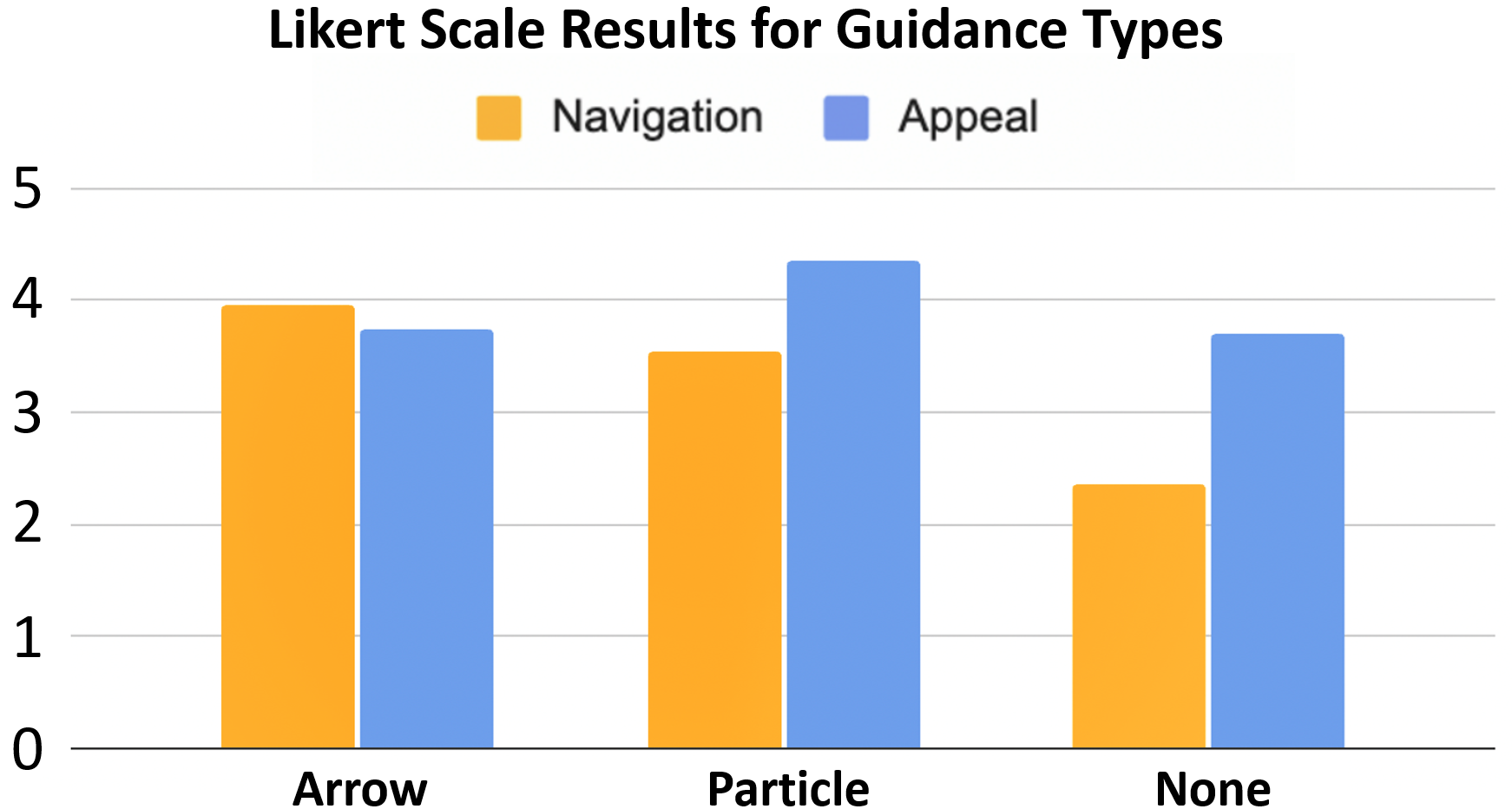}
 \caption{User ratings for the guidance systems. The averages for navigation and appeal, shown here, are based on the users' ratings of the arrow, the particles, and the no guidance conditions. Based on these ratings, users enjoyed the experience (consistent appeal) despite their fluctuating opinions with navigation.}
 \label{fig:indicator-type}
\end{figure}

\subsection{Exploration Time}

The average play time duration per stage was 100.54 seconds, with a perceived time of 165.83 seconds, demonstrating a significant overestimation of time (p-value = $2\times10^{-9}$). Users overestimated the experience duration in 85\% of trials, often perceiving nearly twice the actual duration. Remarkably, users estimated a perceived duration over double the actual time in 41.66\% of cases. These results support previous findings that increased immersion leads to an overestimation of elapsed time \cite{awe, FLAVIAN2019547}.

\subsection{User Evaluation of Guidance Systems}

Figure \ref{fig:indicator-type} shows average user ratings for each guidance system regarding navigation utility and appeal. For navigation, an ANOVA analysis showed significant differences across the arrow group, the particle group, and the no guidance group (F-value = 9.002, p-value $<$ 0.001). Bonferroni post-hoc analysis indicated significant differences between the arrow and no guidance groups (t-value = 3.448, p-value = 0.002), and the no guidance and particle groups (t-value = -3.866, p-value $<$ 0.001). No significant difference was observed between the arrow and particle groups ($t$-value = -0.418, p-value = 1.000).

Another ANOVA analysis categorized by appeal also showed significant differences across the three groups (F-value = 3.509, p-value = 0.037). However, Bonferroni post-hoc analysis did not reveal any significant differences. Between the arrow group and no guidance group, the difference was $t$-value = 0.183, $p$-value = 1.000. Between the arrow group and the particle group, the difference was $t$-value = -2.197, $p$-value = 0.096. Finally, the difference between the no guidance group and the particle group was $t$-value = -2.380, $p$-value = 0.062. These results suggest that both guidance systems had aspects that supported natural user interaction with the performances throughout the various locations in the experiment area.

Our findings confirmed Hypotheses 1 and 2, suggesting users enjoyed the experience and perceived a greater amount of time as having passed. Additionally, the guidance systems proved helpful in locating digital content. However, Hypothesis 3 was not confirmed, as both the particle system and the arrow may have presented distinct advantages.

\subsection{Feedback from the Dancers}

We gathered data from dancers using a 25-question survey, comprising 11 Likert-scale items and 14 open-ended queries.

\subsubsection{Choreographer}
Regarding the use of AR headsets, the choreographer emphasized the importance of allowing the dancers to explore and discover the possibilities and limitations themselves in AR headsets, which is something we did not do enough. A few weeks before the dance session, the choreographer conducted a one-hour wide-scale AR demo to see the virtual stage. This provided them with a better understanding of the augmented reality environment and what could be achieved. Without this first-hand experience, the dancers felt it would have been challenging to contribute to the choreography. The choreographer noted a desire to have a dance studio TV showing storyboards, sketches, and cue sheets so that the dancers and choreographer could maintain the session's flow while enabling them to review and plan their footwork and dance sequence.

\subsubsection{Dancers}

When asked whether background music or virtual stage graphics on big TV screens in the studio would assist in their engagement, the dancers' responses were evenly divided. Three dancers strongly favoring background music, one dancer strongly favoring visual stage imagery, and two dancers preferring visual stage graphics over background music, enabling them to maintain creative engagement in their dance.

The question posed to the dancers regarding the input they received during the dance session was: \textit{``In the dance session, dancers were given input from both the choreographer and the tech specialist. Input from which source was more helpful in allowing you to dance for the imaginary environment?''} Based on the findings, it was revealed that the dancers perceived the instructions from the choreographer as considerably more helpful in facilitating their dance performance for the imaginary environment compared to the input received from the tech specialist. Furthermore, two dancers specifically highlighted the potential benefit of having access to the virtual stage imagery earlier, allowing them adequate time for thoughtful contemplation and strategic planning of their movements within the virtual space.

Despite this, the presence of both the choreographer and tech specialist was deemed necessary by the dancers, as the choreographer assisted in translating the desired mood and tempo for synchronized group performance, while the tech specialist provided the overarching objectives and goals. The dancers expressed appreciation for the provided landscape images, as they greatly aided in creating an imaginary world to dance and choreograph within. 

All six dancers agreed that despite initial concerns about its spatial limitations, the area provided for the project was indeed ideal, with four dancers expressing that it even exceeded their expectations. The dancers acknowledged the value of adapting to spatial limitations, emphasizing the opportunities for creativity and exploration that arise from working within given constraints. Special limitations can actually be beneficial to the choreographic process, presenting an opportunity to explore and play with the dynamics of space. One dancer aptly expressed this sentiment, stating, \textit{``Our job as dancers is to accommodate to our space and choreograph/improvise while working with the given space. The spatial limitations can actually be beneficial to the choreographic process and serve as a creative play of space.”} Furthermore, the inclusion of a padded floor was well-received by the dancers as it offered a balance between sufficient softness for floor movements and adequate traction for more energetic choreography. 

All six dancers expressed their desire to participate in future augmented reality (AR) dance projects, as they found the experience inspiring and believed it positively influenced their own dance practices. One dancer specifically mentioned the unique aspect of focusing solely on the dance itself and being removed from the physical space where the performance would ultimately take place. Another dancer highlighted the joy of creating a new world to dance in and the rewarding nature of the experience. 

\section{Discussion}

The study's purpose was to investigate how audiences consume AR dance performances while walking around in the virtual stage at their own pace. By timing user experience duration in each stage and comparing it with user perception of time, we were able see a noticeable greater perceived passage of time. Detailed pre-and post-study questionnaires also assessed subjective ratings of the experience using each guidance system to lead the integrative narrative. 

\subsection{Impact on Dance Theater}
Our design successfully captivated user engagement. The significant discrepancy between perceived and actual time revealed during the exit interview suggests users' engagement and immersion in the Dynamic Theater system. Using location-based triggers, we created diverse environments and sequentially offered immersive dance experiences within several virtual scenes in a single indoor space.

Our system provides a unique level of interactivity by letting users naturally explore different content zones and interact with virtual performers and the stage, which is not typically enabled by traditional stages. Our AR theater, incorporating the physical layout into the virtual stage design, leveraged the existing infrastructure to create a seamless transition between physical and virtual content. 

\subsection{Use of Guidance System in AR Theater}
The user study data and exit interviews suggest that the particle system, though visually appealing and effective, isn't as intuitive as the arrow for guidance. The arrow had a slight edge in providing clear directions. Its direct, accurate navigation path may be essential for user safety in virtual environments like DreamWalker \cite{yang2019dreamwalker}. However, we aimed to create an experience that compliments both content and usability. In the exit interview, five users expressed difficulty in locating the dancers without any guidance system, confirming that using a guidance system while watching a performance does not diminish enjoyment or appeal when implemented correctly.

\subsection{Supporting Dancers}

Significant creative research supporting and involving actors and dancers as active participants in immersive media has appeared in recent years~\cite{support1, support2, kyan2015approach}. Our AR theater project benefited greatly from the direct contributions of dancers, leading to valuable learning experiences for both the artists and the development team. One noteworthy example is when the dancers proposed a dance routine that involved gesturally creating a tree in response to a virtual forest scene. This approach resonated with the dancers, who expressed a desire for more opportunities to create and explore similar ideas in future opportunities.

Dancers and choreographers emphasized desire for more involvement in the development of immersive technology projects. They highlighted the importance of not underestimating dancers' comprehension abilities. As one choreographer stated, dancers possess problem-solving skills and an intuitive understanding of space and timing. Their ability to perceive subtle movements and changes, as well as envision how the audience will experience the performance and utilize the space and objects, stems from years of training. Therefore, it is crucial to allow dancers to engage with the main technology, such as AR, and provide them with opportunities to explore and take leading roles in these projects.

\section{Conclusion}
Our research fuses wide-area augmented reality, dance, and theater, offering a framework for immersive, interactive dance experiences. We created a location-based theater where users navigate indoor space, interacting with dancers in the augmented stages. Our study emphasizes a guidance system and location-centric approach, curating user movement, engagement, and focus for a seamless and enjoyable experience. Thus, our immersive theater exemplifies an innovative AR application, presenting a novel interactive platform for exploring performing arts and spatial computing.

To overcome hardware limitations and deploy several dancers at once, we employed the billboarding technique. However, abrupt shifts in the user's position can impair spatial awareness, briefly distorting the virtual dancers and the users’ perception of their three-dimensionality. Furthermore, our experiments took place in a corridor-shaped physical environment, which suggests particular walking directions. As such, our findings may be applicable to similar environments, with wider space settings yet to be explored.

We explored location-based AR dance theater, emphasizing guidance systems and content placement to craft curated user experiences. We assessed the guidance system in tandem with our main content and performance as users navigated the space. This system potentially serves as a blueprint for open-world AR applications, recognizing that guidance system selection may vary based on the designers' goals.


\section{Acknowledgments}
This work was supported in part by NSF award IIS-2211784 and UCSB's Media Arts and Technology Program. The authors thank Marko Peljhan, Angela Zhang, and UCSB's RMP program for their support. Furthermore, the authors thank Christina McCarthy for bringing dance performance into our project. Without her support, Dynamic Theater would not exist.


\bibliographystyle{ACM-Reference-Format}
\bibliography{template}


\begin{thebibliography}{54}


\ifx \showCODEN    \undefined \def \showCODEN     #1{\unskip}     \fi
\ifx \showDOI      \undefined \def \showDOI       #1{#1}\fi
\ifx \showISBNx    \undefined \def \showISBNx     #1{\unskip}     \fi
\ifx \showISBNxiii \undefined \def \showISBNxiii  #1{\unskip}     \fi
\ifx \showISSN     \undefined \def \showISSN      #1{\unskip}     \fi
\ifx \showLCCN     \undefined \def \showLCCN      #1{\unskip}     \fi
\ifx \shownote     \undefined \def \shownote      #1{#1}          \fi
\ifx \showarticletitle \undefined \def \showarticletitle #1{#1}   \fi
\ifx \showURL      \undefined \def \showURL       {\relax}        \fi
\providecommand\bibfield[2]{#2}
\providecommand\bibinfo[2]{#2}
\providecommand\natexlab[1]{#1}
\providecommand\showeprint[2][]{arXiv:#2}

\bibitem[Abramovi{\'c}(2020)]%
        {abramovic2020life}
\bibfield{author}{\bibinfo{person}{Marina Abramovi{\'c}}.} \bibinfo{year}{2020}\natexlab{}.
\newblock \bibinfo{title}{The {{Life}}: The World's First {{Mixed Reality}} Performance Artwork - {{Christie}}'s}.
\newblock \bibinfo{howpublished}{https://www.christies.com/features/Marina-Abramovic-The-Life-10193-3.aspx}.
\newblock


\bibitem[Branch et~al\mbox{.}(2021)]%
        {support1}
\bibfield{author}{\bibinfo{person}{Boyd Branch}, \bibinfo{person}{Christos Efstratiou}, \bibinfo{person}{Piotr Mirowski}, \bibinfo{person}{Kory~W. Mathewson}, {and} \bibinfo{person}{Paul Allain}.} \bibinfo{year}{2021}\natexlab{}.
\newblock \showarticletitle{Tele-Immersive Improv: Effects of Immersive Visualisations on Rehearsing and Performing Theatre Online}. In \bibinfo{booktitle}{\emph{Proceedings of the 2021 CHI Conference on Human Factors in Computing Systems}} (Yokohama, Japan) \emph{(\bibinfo{series}{CHI '21})}. \bibinfo{publisher}{Association for Computing Machinery}, \bibinfo{address}{New York, NY, USA}, Article \bibinfo{articleno}{458}, \bibinfo{numpages}{13}~pages.
\newblock
\showISBNx{9781450380966}
\urldef\tempurl%
\url{https://doi.org/10.1145/3411764.3445310}
\showDOI{\tempurl}


\bibitem[Brockhoeft et~al\mbox{.}(2016)]%
        {brockhoeft2016interactive}
\bibfield{author}{\bibinfo{person}{Taylor Brockhoeft}, \bibinfo{person}{Jennifer Petuch}, \bibinfo{person}{James Bach}, \bibinfo{person}{Emil Djerekarov}, \bibinfo{person}{Margareta Ackerman}, {and} \bibinfo{person}{Gary Tyson}.} \bibinfo{year}{2016}\natexlab{}.
\newblock \showarticletitle{Interactive augmented reality for dance}. In \bibinfo{booktitle}{\emph{Proceedings of the Seventh International Conference on Computational Creativity}}. \bibinfo{pages}{396--403}.
\newblock


\bibitem[Buck et~al\mbox{.}(2022)]%
        {buck2022azure}
\bibfield{author}{\bibinfo{person}{Alex Buck}, \bibinfo{person}{Radford Parker}, {and} \bibinfo{person}{Matt Wojciakowski}.} \bibinfo{year}{2022}\natexlab{}.
\newblock \bibinfo{title}{Azure Spatial Anchors overview}.
\newblock \bibinfo{howpublished}{https://learn.microsoft.com/en-us/azure/spatial-anchors/overview}.
\newblock
\newblock
\shownote{Accessed: October 2022}.


\bibitem[Carrigy et~al\mbox{.}(2010)]%
        {location}
\bibfield{author}{\bibinfo{person}{Tara Carrigy}, \bibinfo{person}{Katsiaryna Naliuka}, \bibinfo{person}{Natasa Paterson}, {and} \bibinfo{person}{Mads Haahr}.} \bibinfo{year}{2010}\natexlab{}.
\newblock \showarticletitle{Design and Evaluation of Player Experience of a Location-Based Mobile Game}. In \bibinfo{booktitle}{\emph{Proceedings of the 6th Nordic Conference on Human-Computer Interaction: Extending Boundaries}} (Reykjavik, Iceland) \emph{(\bibinfo{series}{NordiCHI '10})}. \bibinfo{publisher}{Association for Computing Machinery}, \bibinfo{address}{New York, NY, USA}, \bibinfo{pages}{92–101}.
\newblock
\showISBNx{9781605589343}
\urldef\tempurl%
\url{https://doi.org/10.1145/1868914.1868929}
\showDOI{\tempurl}


\bibitem[Cheok et~al\mbox{.}(2004)]%
        {cheok2004human}
\bibfield{author}{\bibinfo{person}{Adrian~David Cheok}, \bibinfo{person}{Kok~Hwee Goh}, \bibinfo{person}{Wei Liu}, \bibinfo{person}{Farzam Farbiz}, \bibinfo{person}{Sze~Lee Teo}, \bibinfo{person}{Hui~Siang Teo}, \bibinfo{person}{Shang~Ping Lee}, \bibinfo{person}{Yu Li}, \bibinfo{person}{Siew~Wan Fong}, {and} \bibinfo{person}{Xubo Yang}.} \bibinfo{year}{2004}\natexlab{}.
\newblock \showarticletitle{Human Pacman: a mobile wide-area entertainment system based on physical, social, and ubiquitous computing}. In \bibinfo{booktitle}{\emph{Proceedings of the 2004 ACM SIGCHI International Conference on Advances in computer entertainment technology}}. \bibinfo{pages}{360--361}.
\newblock


\bibitem[Cheok et~al\mbox{.}(2002)]%
        {cheok2002interactive}
\bibfield{author}{\bibinfo{person}{Adrian~David Cheok}, \bibinfo{person}{Wang Weihua}, \bibinfo{person}{Xubo Yang}, \bibinfo{person}{Simon Prince}, \bibinfo{person}{Fong~Siew Wan}, \bibinfo{person}{Mark Billinghurst}, {and} \bibinfo{person}{Hirokazu Kato}.} \bibinfo{year}{2002}\natexlab{}.
\newblock \showarticletitle{Interactive theatre experience in embodied+ wearable mixed reality space}. In \bibinfo{booktitle}{\emph{Proceedings. International Symposium on Mixed and Augmented Reality}}. IEEE, \bibinfo{pages}{59--317}.
\newblock


\bibitem[Clay et~al\mbox{.}(2012)]%
        {clay2012interactions}
\bibfield{author}{\bibinfo{person}{Alexis Clay}, \bibinfo{person}{Nadine Couture}, \bibinfo{person}{Laurence Nigay}, \bibinfo{person}{Jean-Baptiste De~La~Riviere}, \bibinfo{person}{Jean-Claude Martin}, \bibinfo{person}{Matthieu Courgeon}, \bibinfo{person}{Myriam Desainte-Catherine}, \bibinfo{person}{Emmanuel Orvain}, \bibinfo{person}{Vincent Girondel}, {and} \bibinfo{person}{Ga{\"e}l Domengero}.} \bibinfo{year}{2012}\natexlab{}.
\newblock \showarticletitle{Interactions and systems for augmenting a live dance performance}. In \bibinfo{booktitle}{\emph{2012 IEEE International Symposium on Mixed and Augmented Reality-Arts, Media, and Humanities (ISMAR-AMH)}}. IEEE, \bibinfo{pages}{29--38}.
\newblock


\bibitem[Clay et~al\mbox{.}(2014)]%
        {clay2014integrating}
\bibfield{author}{\bibinfo{person}{Alexis Clay}, \bibinfo{person}{Gaǫl Domenger}, \bibinfo{person}{Julien Conan}, \bibinfo{person}{Axel Domenger}, {and} \bibinfo{person}{Nadine Couture}.} \bibinfo{year}{2014}\natexlab{}.
\newblock \showarticletitle{Integrating augmented reality to enhance expression, interaction \& collaboration in live performances: A ballet dance case study}. In \bibinfo{booktitle}{\emph{2014 IEEE International Symposium on Mixed and Augmented Reality-Media, Art, Social Science, Humanities and Design (ISMAR-MASH'D)}}. IEEE, \bibinfo{pages}{21--29}.
\newblock


\bibitem[Coulombe et~al\mbox{.}(2021)]%
        {coulombe2021virtual}
\bibfield{author}{\bibinfo{person}{Alex Coulombe}, \bibinfo{person}{David Gochfeld}, \bibinfo{person}{Brendan Bradley}, \bibinfo{person}{Kevin Laibson}, \bibinfo{person}{Robert Long}, {and} \bibinfo{person}{Roman Miletitch}.} \bibinfo{year}{2021}\natexlab{}.
\newblock \showarticletitle{Virtual Reality Live Theatre on No Budget: A Model for Independent Theatrical Productions Using Open-Source Social VR}. In \bibinfo{booktitle}{\emph{ACM SIGGRAPH 2021 Educators Forum}} (Virtual Event, USA) \emph{(\bibinfo{series}{SIGGRAPH '21})}. \bibinfo{publisher}{Association for Computing Machinery}, \bibinfo{address}{New York, NY, USA}, Article \bibinfo{articleno}{2}, \bibinfo{numpages}{2}~pages.
\newblock
\showISBNx{9781450383639}
\urldef\tempurl%
\url{https://doi.org/10.1145/3450549.3464413}
\showDOI{\tempurl}


\bibitem[Diana(2018)]%
        {diana2018overlooked}
\bibfield{author}{\bibinfo{person}{Julie Diana}.} \bibinfo{year}{2018}\natexlab{}.
\newblock \bibinfo{title}{The {{Overlooked Art}} of {{Entering}} and {{Exiting}} the {{Stage}} - {{Dance Teacher}}}.
\newblock \bibinfo{howpublished}{https://dance-teacher.com/the-overlooked-art-of-entering-and-exiting-the-stage/}.
\newblock


\bibitem[Dow et~al\mbox{.}(2005)]%
        {dow2005exploring}
\bibfield{author}{\bibinfo{person}{Steven Dow}, \bibinfo{person}{Jaemin Lee}, \bibinfo{person}{Christopher Oezbek}, \bibinfo{person}{Blair Maclntyre}, \bibinfo{person}{Jay~David Bolter}, {and} \bibinfo{person}{Maribeth Gandy}.} \bibinfo{year}{2005}\natexlab{}.
\newblock \showarticletitle{Exploring Spatial Narratives and Mixed Reality Experiences in Oakland Cemetery}. In \bibinfo{booktitle}{\emph{Proceedings of the 2005 ACM SIGCHI International Conference on Advances in Computer Entertainment Technology}} (Valencia, Spain) \emph{(\bibinfo{series}{ACE '05})}. \bibinfo{publisher}{Association for Computing Machinery}, \bibinfo{address}{New York, NY, USA}, \bibinfo{pages}{51–60}.
\newblock
\showISBNx{1595931104}
\urldef\tempurl%
\url{https://doi.org/10.1145/1178477.1178484}
\showDOI{\tempurl}


\bibitem[Erkert(2003)]%
        {erkert2003harnessing}
\bibfield{author}{\bibinfo{person}{Jan Erkert}.} \bibinfo{year}{2003}\natexlab{}.
\newblock \bibinfo{booktitle}{\emph{Harnessing the Wind : The Art of Teaching Modern Dance}}.
\newblock \bibinfo{publisher}{{Champaign, IL : Human Kinetics}}.
\newblock
\showISBNx{978-0-7360-4487-5}


\bibitem[Feiner et~al\mbox{.}(1997)]%
        {feiner1997touring}
\bibfield{author}{\bibinfo{person}{Steven Feiner}, \bibinfo{person}{Blair MacIntyre}, \bibinfo{person}{Tobias H{\"o}llerer}, {and} \bibinfo{person}{Anthony Webster}.} \bibinfo{year}{1997}\natexlab{}.
\newblock \showarticletitle{A touring machine: Prototyping 3D mobile augmented reality systems for exploring the urban environment}.
\newblock \bibinfo{journal}{\emph{Personal Technologies}} \bibinfo{volume}{1}, \bibinfo{number}{4} (\bibinfo{year}{1997}), \bibinfo{pages}{208--217}.
\newblock


\bibitem[Fischer et~al\mbox{.}(2016)]%
        {fischer2016nautilus}
\bibfield{author}{\bibinfo{person}{Andreas Fischer}, \bibinfo{person}{Sara Grimm}, \bibinfo{person}{Valentine Bernasconi}, \bibinfo{person}{Angelika Garz}, \bibinfo{person}{Pascal Buchs}, \bibinfo{person}{Maurizio Caon}, \bibinfo{person}{Omar Abou~Khaled}, \bibinfo{person}{Elena Mugellini}, \bibinfo{person}{Franziska Meyer}, {and} \bibinfo{person}{Claudia Wagner}.} \bibinfo{year}{2016}\natexlab{}.
\newblock \showarticletitle{Nautilus: Real-Time Interaction Between Dancers and Augmented Reality with Pixel-Cloud Avatars}. In \bibinfo{booktitle}{\emph{28i{\`e}me conf{\'e}rence francophone sur l’Interaction Homme-Machine}}. \bibinfo{pages}{50--57}.
\newblock


\bibitem[Flavián et~al\mbox{.}(2019)]%
        {FLAVIAN2019547}
\bibfield{author}{\bibinfo{person}{Carlos Flavián}, \bibinfo{person}{Sergio Ibáñez-Sánchez}, {and} \bibinfo{person}{Carlos Orús}.} \bibinfo{year}{2019}\natexlab{}.
\newblock \showarticletitle{The impact of virtual, augmented and mixed reality technologies on the customer experience}.
\newblock \bibinfo{journal}{\emph{Journal of Business Research}}  \bibinfo{volume}{100} (\bibinfo{year}{2019}), \bibinfo{pages}{547--560}.
\newblock
\showISSN{0148-2963}
\urldef\tempurl%
\url{https://doi.org/10.1016/j.jbusres.2018.10.050}
\showDOI{\tempurl}


\bibitem[Fujihata(2022)]%
        {fujihata2022behere}
\bibfield{author}{\bibinfo{person}{Masaki Fujihata}.} \bibinfo{year}{2022}\natexlab{}.
\newblock \bibinfo{title}{{{BeHere}} 1942, {{Japanese American National Museum}}}.
\newblock \bibinfo{howpublished}{https://www.janm.org/exhibits/behere1942}.
\newblock


\bibitem[Geigel and Schweppe(2004)]%
        {geigel2004theatrical}
\bibfield{author}{\bibinfo{person}{Joe Geigel} {and} \bibinfo{person}{Marla Schweppe}.} \bibinfo{year}{2004}\natexlab{}.
\newblock \showarticletitle{Theatrical Storytelling in a Virtual Space}. In \bibinfo{booktitle}{\emph{Proceedings of the 1st ACM Workshop on Story Representation, Mechanism and Context}} (New York, NY, USA) \emph{(\bibinfo{series}{SRMC '04})}. \bibinfo{publisher}{Association for Computing Machinery}, \bibinfo{address}{New York, NY, USA}, \bibinfo{pages}{39–46}.
\newblock
\showISBNx{1581139314}
\urldef\tempurl%
\url{https://doi.org/10.1145/1026633.1026642}
\showDOI{\tempurl}


\bibitem[Geigel et~al\mbox{.}(2020)]%
        {geigel2020digital}
\bibfield{author}{\bibinfo{person}{Joe Geigel}, \bibinfo{person}{Kunal~Shailesh Shitut}, \bibinfo{person}{Juilee Decker}, \bibinfo{person}{Amanda Doherty}, {and} \bibinfo{person}{Gary Jacobs}.} \bibinfo{year}{2020}\natexlab{}.
\newblock \showarticletitle{The Digital Docent: XR Storytelling for a Living History Museum}. In \bibinfo{booktitle}{\emph{Proceedings of the 26th ACM Symposium on Virtual Reality Software and Technology}} (Virtual Event, Canada) \emph{(\bibinfo{series}{VRST '20})}. \bibinfo{publisher}{Association for Computing Machinery}, \bibinfo{address}{New York, NY, USA}, Article \bibinfo{articleno}{74}, \bibinfo{numpages}{3}~pages.
\newblock
\showISBNx{9781450376198}
\urldef\tempurl%
\url{https://doi.org/10.1145/3385956.3422090}
\showDOI{\tempurl}


\bibitem[Gochfeld et~al\mbox{.}(2018)]%
        {gochfeld2018holojam}
\bibfield{author}{\bibinfo{person}{David Gochfeld}, \bibinfo{person}{Corinne Brenner}, \bibinfo{person}{Kris Layng}, \bibinfo{person}{Sebastian Herscher}, \bibinfo{person}{Connor DeFanti}, \bibinfo{person}{Marta Olko}, \bibinfo{person}{David Shinn}, \bibinfo{person}{Stephanie Riggs}, \bibinfo{person}{Clara Fern{\'a}ndez-Vara}, {and} \bibinfo{person}{Ken Perlin}.} \bibinfo{year}{2018}\natexlab{}.
\newblock \showarticletitle{Holojam in wonderland: immersive mixed reality theater}.
\newblock \bibinfo{journal}{\emph{Leonardo}} \bibinfo{volume}{51}, \bibinfo{number}{4} (\bibinfo{year}{2018}), \bibinfo{pages}{362--367}.
\newblock


\bibitem[Gochfeld et~al\mbox{.}(2022)]%
        {gochfeld2022tale}
\bibfield{author}{\bibinfo{person}{David Gochfeld}, \bibinfo{person}{Alex Coulombe}, \bibinfo{person}{Yu-Jun Yeh}, \bibinfo{person}{Robert Lester}, \bibinfo{person}{Robert~Barry Fleming}, \bibinfo{person}{Zachary Meicher-Buzzi}, {and} \bibinfo{person}{Ari Tarr}.} \bibinfo{year}{2022}\natexlab{}.
\newblock \showarticletitle{A Tale of Two Productions: A Christmas Carol On Stage and in VR}.
\newblock \bibinfo{journal}{\emph{Proc. ACM Comput. Graph. Interact. Tech.}} \bibinfo{volume}{5}, \bibinfo{number}{4}, Article \bibinfo{articleno}{41} (\bibinfo{date}{sep} \bibinfo{year}{2022}), \bibinfo{numpages}{9}~pages.
\newblock
\urldef\tempurl%
\url{https://doi.org/10.1145/3533612}
\showDOI{\tempurl}


\bibitem[Han et~al\mbox{.}(2023)]%
        {han2023architectural}
\bibfield{author}{\bibinfo{person}{Jihae Han}, \bibinfo{person}{Andrew Vande~Moere}, {and} \bibinfo{person}{Adalberto~L. Simeone}.} \bibinfo{year}{2023}\natexlab{}.
\newblock \showarticletitle{Architectural Narrative VR: Towards Generatively Designing Natural Walkable Spaces}. In \bibinfo{booktitle}{\emph{Proceedings of the 2023 ACM Designing Interactive Systems Conference}} (Pittsburgh, PA, USA) \emph{(\bibinfo{series}{DIS '23})}. \bibinfo{publisher}{Association for Computing Machinery}, \bibinfo{address}{New York, NY, USA}, \bibinfo{pages}{523–536}.
\newblock
\showISBNx{9781450398930}
\urldef\tempurl%
\url{https://doi.org/10.1145/3563657.3596008}
\showDOI{\tempurl}


\bibitem[Herscher et~al\mbox{.}(2019)]%
        {herscher2019cavrn}
\bibfield{author}{\bibinfo{person}{Sebastian Herscher}, \bibinfo{person}{Connor DeFanti}, \bibinfo{person}{Nicholas~Gregory Vitovitch}, \bibinfo{person}{Corinne Brenner}, \bibinfo{person}{Haijun Xia}, \bibinfo{person}{Kris Layng}, {and} \bibinfo{person}{Ken Perlin}.} \bibinfo{year}{2019}\natexlab{}.
\newblock \showarticletitle{CAVRN: an exploration and evaluation of a collective audience virtual reality nexus experience}. In \bibinfo{booktitle}{\emph{Proceedings of the 32nd Annual ACM Symposium on User Interface Software and Technology}}. \bibinfo{pages}{1137--1150}.
\newblock


\bibitem[Kim et~al\mbox{.}(2022)]%
        {kim2022investigating}
\bibfield{author}{\bibinfo{person}{You-Jin Kim}, \bibinfo{person}{Radha Kumaran}, \bibinfo{person}{Ehsan Sayyad}, \bibinfo{person}{Anne Milner}, \bibinfo{person}{Tom Bullock}, \bibinfo{person}{Barry Giesbrecht}, {and} \bibinfo{person}{Tobias Hollerer}.} \bibinfo{year}{2022}\natexlab{}.
\newblock \showarticletitle{Investigating Search Among Physical and Virtual Objects Under Different Lighting Conditions}.
\newblock \bibinfo{journal}{\emph{IEEE Transactions on Visualization and Computer Graphics}} (\bibinfo{year}{2022}), \bibinfo{pages}{1--11}.
\newblock
\showISSN{1077-2626}
\urldef\tempurl%
\url{https://doi.org/10.1109/TVCG.2022.3203093}
\showDOI{\tempurl}


\bibitem[Kumaran et~al\mbox{.}(2023)]%
        {kumaran2023impact}
\bibfield{author}{\bibinfo{person}{Radha Kumaran}, \bibinfo{person}{You-Jin Kim}, \bibinfo{person}{Anne~E Milner}, \bibinfo{person}{Tom Bullock}, \bibinfo{person}{Barry Giesbrecht}, {and} \bibinfo{person}{Tobias H\"{o}llerer}.} \bibinfo{year}{2023}\natexlab{}.
\newblock \showarticletitle{The Impact of Navigation Aids on Search Performance and Object Recall in Wide-Area Augmented Reality}. In \bibinfo{booktitle}{\emph{Proceedings of the 2023 CHI Conference on Human Factors in Computing Systems}} (Hamburg, Germany) \emph{(\bibinfo{series}{CHI '23})}. \bibinfo{publisher}{Association for Computing Machinery}, \bibinfo{address}{New York, NY, USA}, Article \bibinfo{articleno}{710}, \bibinfo{numpages}{17}~pages.
\newblock
\showISBNx{9781450394215}
\urldef\tempurl%
\url{https://doi.org/10.1145/3544548.3581413}
\showDOI{\tempurl}


\bibitem[Kyan et~al\mbox{.}(2015)]%
        {kyan2015approach}
\bibfield{author}{\bibinfo{person}{Matthew Kyan}, \bibinfo{person}{Guoyu Sun}, \bibinfo{person}{Haiyan Li}, \bibinfo{person}{Ling Zhong}, \bibinfo{person}{Paisarn Muneesawang}, \bibinfo{person}{Nan Dong}, \bibinfo{person}{Bruce Elder}, {and} \bibinfo{person}{Ling Guan}.} \bibinfo{year}{2015}\natexlab{}.
\newblock \showarticletitle{An approach to ballet dance training through ms kinect and visualization in a cave virtual reality environment}.
\newblock \bibinfo{journal}{\emph{ACM Transactions on Intelligent Systems and Technology (TIST)}} \bibinfo{volume}{6}, \bibinfo{number}{2} (\bibinfo{year}{2015}), \bibinfo{pages}{1--37}.
\newblock


\bibitem[Layng et~al\mbox{.}(2019)]%
        {layng2019cave}
\bibfield{author}{\bibinfo{person}{Kris Layng}, \bibinfo{person}{Ken Perlin}, \bibinfo{person}{Sebastian Herscher}, \bibinfo{person}{Corinne Brenner}, {and} \bibinfo{person}{Thomas Meduri}.} \bibinfo{year}{2019}\natexlab{}.
\newblock \showarticletitle{Cave: making collective virtual narrative}.
\newblock In \bibinfo{booktitle}{\emph{ACM SIGGRAPH 2019 Art Gallery}}. \bibinfo{pages}{1--8}.
\newblock


\bibitem[Lehto et~al\mbox{.}(2020)]%
        {lehto2020augmented}
\bibfield{author}{\bibinfo{person}{Anttoni Lehto}, \bibinfo{person}{Nina Luostarinen}, {and} \bibinfo{person}{Paula Kostia}.} \bibinfo{year}{2020}\natexlab{}.
\newblock \showarticletitle{Augmented reality gaming as a tool for subjectivizing visitor experience at cultural heritage locations—case lights on!}
\newblock \bibinfo{journal}{\emph{Journal on Computing and Cultural Heritage (JOCCH)}} \bibinfo{volume}{13}, \bibinfo{number}{4} (\bibinfo{year}{2020}), \bibinfo{pages}{1--16}.
\newblock


\bibitem[Lella and Lipsman(2014)]%
        {lella2014us}
\bibfield{author}{\bibinfo{person}{Adam Lella} {and} \bibinfo{person}{Andrew Lipsman}.} \bibinfo{year}{2014}\natexlab{}.
\newblock \showarticletitle{The US mobile app report}.
\newblock \bibinfo{howpublished}{https://www.comscore.com/Insights/Presentations-and-Whitepapers/2014/The-US-Mobile-App-Report}.
\newblock \bibinfo{journal}{\emph{Tech. Rep.}}  \bibinfo{volume}{8} (\bibinfo{year}{2014}).
\newblock


\bibitem[Li et~al\mbox{.}(2023)]%
        {li2023locationaware}
\bibfield{author}{\bibinfo{person}{Wanwan Li}, \bibinfo{person}{Changyang Li}, \bibinfo{person}{Minyoung Kim}, \bibinfo{person}{Haikun Huang}, {and} \bibinfo{person}{Lap-Fai Yu}.} \bibinfo{year}{2023}\natexlab{}.
\newblock \showarticletitle{Location-Aware Adaptation of Augmented Reality Narratives}. In \bibinfo{booktitle}{\emph{Proceedings of the 2023 CHI Conference on Human Factors in Computing Systems}} (Hamburg, Germany) \emph{(\bibinfo{series}{CHI '23})}. \bibinfo{publisher}{Association for Computing Machinery}, \bibinfo{address}{New York, NY, USA}, Article \bibinfo{articleno}{33}, \bibinfo{numpages}{15}~pages.
\newblock
\showISBNx{9781450394215}
\urldef\tempurl%
\url{https://doi.org/10.1145/3544548.3580978}
\showDOI{\tempurl}


\bibitem[Lyons et~al\mbox{.}(2023)]%
        {lyons2023gumball}
\bibfield{author}{\bibinfo{person}{Deirdre~V. Lyons}, \bibinfo{person}{Christopher~Lane Davis}, \bibinfo{person}{Stephen Butchko}, \bibinfo{person}{Whitton Frank}, \bibinfo{person}{Brian Tull}, {and} \bibinfo{person}{Braden Roy}.} \bibinfo{year}{2023}\natexlab{}.
\newblock \showarticletitle{Gumball Dreams: Live Theatre in VR}. In \bibinfo{booktitle}{\emph{ACM SIGGRAPH 2023 Immersive Pavilion}} (Los Angeles, CA, USA) \emph{(\bibinfo{series}{SIGGRAPH '23})}. \bibinfo{publisher}{Association for Computing Machinery}, \bibinfo{address}{New York, NY, USA}, Article \bibinfo{articleno}{8}, \bibinfo{numpages}{2}~pages.
\newblock
\showISBNx{9798400701511}
\urldef\tempurl%
\url{https://doi.org/10.1145/3588027.3595593}
\showDOI{\tempurl}


\bibitem[Madsen et~al\mbox{.}(2022)]%
        {madsen2022fear}
\bibfield{author}{\bibinfo{person}{Peter Madsen}, \bibinfo{person}{Henning Pohl}, {and} \bibinfo{person}{Timothy Merritt}.} \bibinfo{year}{2022}\natexlab{}.
\newblock \showarticletitle{Fear Inducing Play in an AR Escape Room with Human and Robotic NPCs}. In \bibinfo{booktitle}{\emph{Extended Abstracts of the 2022 Annual Symposium on Computer-Human Interaction in Play}}. \bibinfo{pages}{38--43}.
\newblock


\bibitem[McAuley(1999)]%
        {mcauley1999space}
\bibfield{author}{\bibinfo{person}{Gay McAuley}.} \bibinfo{year}{1999}\natexlab{}.
\newblock \bibinfo{booktitle}{\emph{Space in {{Performance}}: {{Making Meaning}} in the {{Theatre}}}}.
\newblock \bibinfo{publisher}{{University of Michigan Press}}.
\newblock
\showISBNx{978-0-472-11004-9}


\bibitem[Mileva(2021)]%
        {ARGulliver}
\bibfield{author}{\bibinfo{person}{Gergana Mileva}.} \bibinfo{year}{2021}\natexlab{}.
\newblock \bibinfo{title}{Traditional Theater Gets an Augmented Reality Makeover}.
\newblock
\newblock


\bibitem[Morrison et~al\mbox{.}(2009)]%
        {morrison2009like}
\bibfield{author}{\bibinfo{person}{Ann Morrison}, \bibinfo{person}{Antti Oulasvirta}, \bibinfo{person}{Peter Peltonen}, \bibinfo{person}{Saija Lemmela}, \bibinfo{person}{Giulio Jacucci}, \bibinfo{person}{Gerhard Reitmayr}, \bibinfo{person}{Jaana N{\"a}s{\"a}nen}, {and} \bibinfo{person}{Antti Juustila}.} \bibinfo{year}{2009}\natexlab{}.
\newblock \showarticletitle{Like bees around the hive: a comparative study of a mobile augmented reality map}. In \bibinfo{booktitle}{\emph{Proceedings of the SIGCHI conference on human factors in computing systems}}. \bibinfo{pages}{1889--1898}.
\newblock


\bibitem[Nicholas et~al\mbox{.}(2021)]%
        {nicholas2021expanding}
\bibfield{author}{\bibinfo{person}{Molly~Jane Nicholas}, \bibinfo{person}{Stephanie~Claudino Daffara}, {and} \bibinfo{person}{Eric Paulos}.} \bibinfo{year}{2021}\natexlab{}.
\newblock \showarticletitle{Expanding the Design Space for Technology-Mediated Theatre Experiences}. In \bibinfo{booktitle}{\emph{Designing Interactive Systems Conference 2021}}. \bibinfo{pages}{2026--2038}.
\newblock


\bibitem[{Oculus Story Studio}(2017)]%
        {oculus2017dear}
\bibfield{author}{\bibinfo{person}{{Oculus Story Studio}}.} \bibinfo{year}{2017}\natexlab{}.
\newblock \bibinfo{booktitle}{\emph{Dear Angelica}}.
\newblock
\newblock
\shownote{https://www.oculus.com/experiences/rift/1174445049267874/}.


\bibitem[Pietroszek et~al\mbox{.}(2022a)]%
        {pietroszek2022dill}
\bibfield{author}{\bibinfo{person}{Krzysztof Pietroszek}, \bibinfo{person}{Manuel Rebol}, {and} \bibinfo{person}{Becky Lake}.} \bibinfo{year}{2022}\natexlab{a}.
\newblock \showarticletitle{Dill Pickle: Interactive Theatre Play in Virtual Reality}. In \bibinfo{booktitle}{\emph{Proceedings of the 28th ACM Symposium on Virtual Reality Software and Technology}}. \bibinfo{pages}{1--2}.
\newblock


\bibitem[Pietroszek et~al\mbox{.}(2022b)]%
        {pietroszek2022meeting}
\bibfield{author}{\bibinfo{person}{Krzysztof Pietroszek}, \bibinfo{person}{Manuel Rebol}, {and} \bibinfo{person}{Becky Lake}.} \bibinfo{year}{2022}\natexlab{b}.
\newblock \showarticletitle{The Meeting: Volumetric Participatory theatre Play in Mixed Reality}. In \bibinfo{booktitle}{\emph{Proceedings of the 10th International Conference on Human-Agent Interaction}}. \bibinfo{pages}{330--332}.
\newblock


\bibitem[Ppali et~al\mbox{.}(2022)]%
        {support2}
\bibfield{author}{\bibinfo{person}{Sophia Ppali}, \bibinfo{person}{Vali Lalioti}, \bibinfo{person}{Boyd Branch}, \bibinfo{person}{Chee~Siang Ang}, \bibinfo{person}{Andrew~J. Thomas}, \bibinfo{person}{Bea~S. Wohl}, {and} \bibinfo{person}{Alexandra Covaci}.} \bibinfo{year}{2022}\natexlab{}.
\newblock \showarticletitle{Keep the VRhythm Going: A Musician-Centred Study Investigating How Virtual Reality Can Support Creative Musical Practice}. In \bibinfo{booktitle}{\emph{Proceedings of the 2022 CHI Conference on Human Factors in Computing Systems}} (New Orleans, LA, USA) \emph{(\bibinfo{series}{CHI '22})}. \bibinfo{publisher}{Association for Computing Machinery}, \bibinfo{address}{New York, NY, USA}, Article \bibinfo{articleno}{220}, \bibinfo{numpages}{19}~pages.
\newblock
\showISBNx{9781450391573}
\urldef\tempurl%
\url{https://doi.org/10.1145/3491102.3501922}
\showDOI{\tempurl}


\bibitem[Rompapas et~al\mbox{.}(2018)]%
        {rompapas2018holoroyale}
\bibfield{author}{\bibinfo{person}{Damien Rompapas}, \bibinfo{person}{Christian Sandor}, \bibinfo{person}{Alexander Plopski}, \bibinfo{person}{Daniel Saakes}, \bibinfo{person}{Dong~Hyeok Yun}, \bibinfo{person}{Takafumi Taketomi}, {and} \bibinfo{person}{Hirokazu Kato}.} \bibinfo{year}{2018}\natexlab{}.
\newblock \showarticletitle{Holoroyale: A large scale high fidelity augmented reality game}. In \bibinfo{booktitle}{\emph{Adjunct Proceedings of the 31st Annual ACM Symposium on User Interface Software and Technology}}. \bibinfo{pages}{163--165}.
\newblock


\bibitem[Rompapas et~al\mbox{.}(2019)]%
        {rompapas2019towards}
\bibfield{author}{\bibinfo{person}{Damien~C. Rompapas}, \bibinfo{person}{Christian Sandor}, \bibinfo{person}{Alexander Plopski}, \bibinfo{person}{Daniel Saakes}, \bibinfo{person}{Joongi Shin}, \bibinfo{person}{Takafumi Taketomi}, {and} \bibinfo{person}{Hirokazu Kato}.} \bibinfo{year}{2019}\natexlab{}.
\newblock \showarticletitle{Towards large scale high fidelity collaborative augmented reality}.
\newblock \bibinfo{journal}{\emph{Computers \& Graphics}}  \bibinfo{volume}{84} (\bibinfo{year}{2019}), \bibinfo{pages}{24--41}.
\newblock


\bibitem[Rudd et~al\mbox{.}(2012)]%
        {awe}
\bibfield{author}{\bibinfo{person}{Melanie Rudd}, \bibinfo{person}{Kathleen~D. Vohs}, {and} \bibinfo{person}{Jennifer Aaker}.} \bibinfo{year}{2012}\natexlab{}.
\newblock \showarticletitle{Awe Expands People’s Perception of Time, Alters Decision Making, and Enhances Well-Being}.
\newblock \bibinfo{journal}{\emph{Psychological Science}} \bibinfo{volume}{23}, \bibinfo{number}{10} (\bibinfo{year}{2012}), \bibinfo{pages}{1130--1136}.
\newblock
\urldef\tempurl%
\url{https://doi.org/10.1177/0956797612438731}
\showDOI{\tempurl}
\showeprint{https://doi.org/10.1177/0956797612438731}
\newblock
\shownote{PMID: 22886132}.


\bibitem[Sayyad et~al\mbox{.}(2020)]%
        {sayyad2020walking}
\bibfield{author}{\bibinfo{person}{Ehsan Sayyad}, \bibinfo{person}{Misha Sra}, {and} \bibinfo{person}{Tobias H{\"o}llerer}.} \bibinfo{year}{2020}\natexlab{}.
\newblock \showarticletitle{Walking and teleportation in wide-area virtual reality experiences}. In \bibinfo{booktitle}{\emph{2020 IEEE International Symposium on Mixed and Augmented Reality (ISMAR)}}. IEEE, \bibinfo{pages}{608--617}.
\newblock


\bibitem[Singh et~al\mbox{.}(2021)]%
        {singh2021story}
\bibfield{author}{\bibinfo{person}{Abbey Singh}, \bibinfo{person}{Ramanpreet Kaur}, \bibinfo{person}{Peter Haltner}, \bibinfo{person}{Matthew Peachey}, \bibinfo{person}{Mar Gonzalez-Franco}, \bibinfo{person}{Joseph Malloch}, {and} \bibinfo{person}{Derek Reilly}.} \bibinfo{year}{2021}\natexlab{}.
\newblock \showarticletitle{Story creatar: a toolkit for spatially-adaptive augmented reality storytelling}. In \bibinfo{booktitle}{\emph{2021 IEEE Virtual Reality and 3D User Interfaces (VR)}}. IEEE, \bibinfo{pages}{713--722}.
\newblock


\bibitem[Tao et~al\mbox{.}(2023)]%
        {self}
\bibfield{author}{\bibinfo{person}{Yujie Tao}, \bibinfo{person}{Cheng~Yao Wang}, \bibinfo{person}{Andrew~D Wilson}, \bibinfo{person}{Eyal Ofek}, {and} \bibinfo{person}{Mar Gonzalez-Franco}.} \bibinfo{year}{2023}\natexlab{}.
\newblock \showarticletitle{Embodying Physics-Aware Avatars in Virtual Reality}. In \bibinfo{booktitle}{\emph{Proceedings of the 2023 CHI Conference on Human Factors in Computing Systems}} (Hamburg, Germany) \emph{(\bibinfo{series}{CHI '23})}. \bibinfo{publisher}{Association for Computing Machinery}, \bibinfo{address}{New York, NY, USA}, Article \bibinfo{articleno}{254}, \bibinfo{numpages}{15}~pages.
\newblock
\showISBNx{9781450394215}
\urldef\tempurl%
\url{https://doi.org/10.1145/3544548.3580979}
\showDOI{\tempurl}


\bibitem[{Tender Claws}(2019)]%
        {tender2019under}
\bibfield{author}{\bibinfo{person}{{Tender Claws}}.} \bibinfo{year}{2019}\natexlab{}.
\newblock \bibinfo{booktitle}{\emph{The Under Presents}}.
\newblock
\newblock
\shownote{https://www.oculus.com/experiences/quest/1917371471713228/}.


\bibitem[Thomas et~al\mbox{.}(2002)]%
        {thomas2002first}
\bibfield{author}{\bibinfo{person}{Bruce Thomas}, \bibinfo{person}{Ben Close}, \bibinfo{person}{John Donoghue}, \bibinfo{person}{John Squires}, \bibinfo{person}{Phillip~De Bondi}, {and} \bibinfo{person}{Wayne Piekarski}.} \bibinfo{year}{2002}\natexlab{}.
\newblock \showarticletitle{First person indoor/outdoor augmented reality application: ARQuake}.
\newblock \bibinfo{journal}{\emph{Personal and Ubiquitous Computing}} \bibinfo{volume}{6}, \bibinfo{number}{1} (\bibinfo{year}{2002}), \bibinfo{pages}{75--86}.
\newblock


\bibitem[Thomas et~al\mbox{.}(2000)]%
        {thomas2000arquake}
\bibfield{author}{\bibinfo{person}{Bruce Thomas}, \bibinfo{person}{Benjamin Close}, \bibinfo{person}{John Donoghue}, \bibinfo{person}{John Squires}, \bibinfo{person}{Phillip De~Bondi}, \bibinfo{person}{Michael Morris}, {and} \bibinfo{person}{Wayne Piekarski}.} \bibinfo{year}{2000}\natexlab{}.
\newblock \showarticletitle{ARQuake: An outdoor/indoor augmented reality first person application}. In \bibinfo{booktitle}{\emph{Digest of papers. Fourth international symposium on wearable computers}}. IEEE, \bibinfo{pages}{139--146}.
\newblock


\bibitem[Weng et~al\mbox{.}(2011)]%
        {weng2011soul}
\bibfield{author}{\bibinfo{person}{DongDong Weng}, \bibinfo{person}{WeiPeng Xu}, \bibinfo{person}{Dong Li}, \bibinfo{person}{YongTian Wang}, {and} \bibinfo{person}{Yue Liu}.} \bibinfo{year}{2011}\natexlab{}.
\newblock \showarticletitle{“Soul Hunter”: A novel augmented reality application in theme parks}. In \bibinfo{booktitle}{\emph{2011 10th IEEE International Symposium on Mixed and Augmented Reality}}. IEEE, \bibinfo{pages}{279--280}.
\newblock


\bibitem[Windschitl and Winn(2000)]%
        {windschitl2000virtual}
\bibfield{author}{\bibinfo{person}{Mark Windschitl} {and} \bibinfo{person}{Bill Winn}.} \bibinfo{year}{2000}\natexlab{}.
\newblock \showarticletitle{A virtual environment designed to help students understand science}. In \bibinfo{booktitle}{\emph{International Conference of the Learning Sciences}}. Psychology Press, \bibinfo{pages}{302--308}.
\newblock


\bibitem[Winn(2001)]%
        {winn2001learning}
\bibfield{author}{\bibinfo{person}{Bill Winn}.} \bibinfo{year}{2001}\natexlab{}.
\newblock \showarticletitle{Learning through virtual reality}.
\newblock \bibinfo{journal}{\emph{Retrieved on February}}  \bibinfo{volume}{23} (\bibinfo{year}{2001}), \bibinfo{pages}{2009}.
\newblock


\bibitem[Wither et~al\mbox{.}(2010)]%
        {wither2010westwood}
\bibfield{author}{\bibinfo{person}{Jason Wither}, \bibinfo{person}{Rebecca Allen}, \bibinfo{person}{Vids Samanta}, \bibinfo{person}{Juha Hemanus}, \bibinfo{person}{Yun-Ta Tsai}, \bibinfo{person}{Ronald Azuma}, \bibinfo{person}{Will Carter}, \bibinfo{person}{Rachel Hinman}, {and} \bibinfo{person}{Thommen Korah}.} \bibinfo{year}{2010}\natexlab{}.
\newblock \showarticletitle{The westwood experience: connecting story to locations via mixed reality}. In \bibinfo{booktitle}{\emph{2010 IEEE International Symposium on Mixed and Augmented Reality-Arts, Media, and Humanities}}. IEEE, \bibinfo{pages}{39--46}.
\newblock


\bibitem[Yang et~al\mbox{.}(2019)]%
        {yang2019dreamwalker}
\bibfield{author}{\bibinfo{person}{Jackie Yang}, \bibinfo{person}{Christian Holz}, \bibinfo{person}{Eyal Ofek}, {and} \bibinfo{person}{Andrew~D Wilson}.} \bibinfo{year}{2019}\natexlab{}.
\newblock \showarticletitle{Dreamwalker: Substituting real-world walking experiences with a virtual reality}. In \bibinfo{booktitle}{\emph{Proceedings of the 32nd Annual ACM Symposium on User Interface Software and Technology}}. \bibinfo{pages}{1093--1107}.
\newblock


\end{thebibliography}

\end{document}